\documentclass[aps,prl,twocolumn,superscriptaddress,10pt,article,showpacs,longbibliography]{revtex4-2}
\usepackage[utf8]{inputenc}
\usepackage{amsfonts}
\usepackage{amsmath}
\usepackage{amsthm}
\usepackage{braket}
\usepackage{graphicx}
\usepackage{hyperref}
\hypersetup{
    colorlinks=true,
    linkcolor=blue,
    filecolor=cyan,      
    urlcolor=magenta,
    citecolor=purple,
}

\usepackage[dvipsnames]{xcolor}
\usepackage{amssymb}
\usepackage{dsfont}
\usepackage{makecell}
\usepackage{verbatim}
\usepackage{amsmath,amscd}
\usepackage{multirow}
\usepackage[normalem]{ulem}
\usepackage{bm}

\def \q{{\mathbf q}}

\def \n{{\mathbf n}}

\def \r{{\mathbf r}}

\def \B{{\mathbf B}}

\def \S{{\mathbf S}}

\def \N{{\mathcal N}}

\def \beq{\begin{eqnarray}}
\def \eeq{\end{eqnarray}}

\begin{document}

\title{Quantum noise spectroscopy of dynamical critical phenomena}

\author{Francisco Machado}
\affiliation{ITAMP, Harvard-Smithsonian Center for Astrophysics, Cambridge, Massachusetts, 02138, USA}
\affiliation{Department of Physics, Harvard University, MA 02138, USA}
\affiliation{Department of Physics, University of California, Berkeley, CA 94720, USA}

\author{Eugene A. Demler}
\affiliation{Institute for Theoretical Physics, ETH Zurich, 8093 Zurich, Switzerland}

\author{Norman Y. Yao}
\affiliation{Department of Physics, Harvard University, MA 02138, USA}
\affiliation{Department of Physics, University of California, Berkeley, CA 94720, USA}

\author{Shubhayu Chatterjee}
\affiliation{Department of Physics, University of California, Berkeley, CA 94720, USA}
\affiliation{Department of Physics, Carnegie Mellon University, Pittsburgh, PA 15213, USA}

\begin{abstract}
The transition between distinct phases of matter is characterized by the nature of fluctuations near the critical point. 
%
%
We demonstrate that noise spectroscopy can not only diagnose the presence of a phase transition, but can also determine fundamental properties of its criticality. 
In particular, by analyzing a scaling collapse of the  decoherence profile, one can directly  extract the critical exponents of the transition and identify its universality class. 
Our approach naturally captures the presence of conservation laws and distinguishes between classical and quantum phase transitions.
In the context of quantum magnetism, our proposal complements existing techniques and provides a novel toolset optimized for interrogating two-dimensional magnetic materials.
\end{abstract}

\maketitle

Continuous phase transitions exhibit remarkable universality across disparate physical systems~\cite{landau2013statistical,CardyBook,goldenfeld2018lectures,coleman_2015,ssbook,SondhiQPT,vojta2000quantum,vojta2003quantum,HH77}.
Owing to the complex interplay between charge, spin and lattice degrees of freedom, quantum materials have emerged as a particularly fruitful setting for exploring phase transitions~\cite{keimer2017physics,basov2017towards,li2021phase,2021QMRoadmap}. 
To fully characterize such critical phenomena, one must accurately measure both \emph{static} and \emph{dynamical} correlations.
The diverging length and time scales at  phase transitions require the ability to simultaneously probe low energies and momenta.
To obtain such data, one often resorts to one of two broad classes of experimental probes: 
scattering techniques, where  correlations in the material lead to momentum and energy shifts on scattered particles (e.g.~neutron scattering, Brillouin light scattering, magnetic optical Kerr, etc.)~\cite{VanHove1954,AdvancesBLS,haider2017reviewMOKE, price2013Neutron,tan:2019,zhou:2021};
and magnetic resonance techniques, where fluctuations in the material generate frequency shifts on localized probe spins~(e.g.~$\mu$SR, ESR, NMR, etc.)~\cite{hore2015NMR,berthier2017NMR,katsumata2000ESR, hillier2022muon}.

\begin{figure}[h!]
    \centering
    \includegraphics[trim={0 0 2mm 0},clip,width = 3.4in]{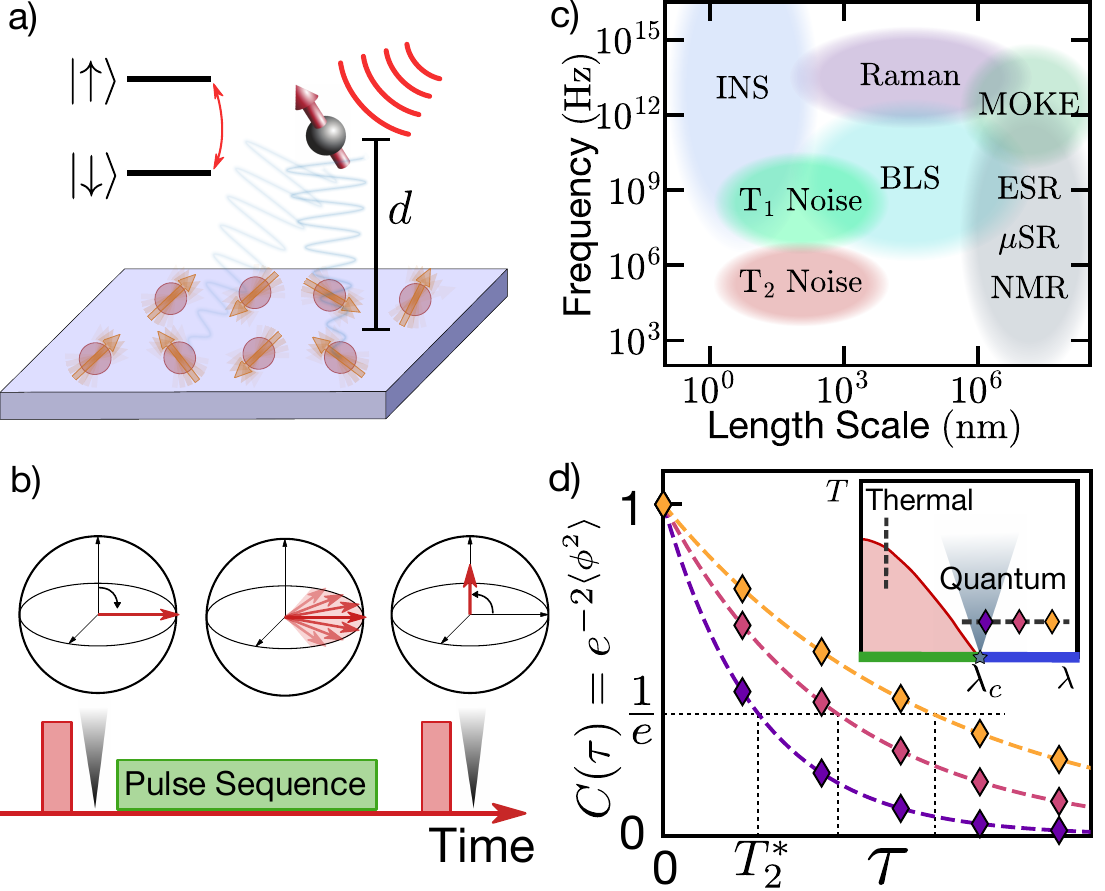}
    \caption{
    {\bf a)} Schematic of the setup: a two-level probe qubit or qubit ensemble is placed a  distance $d$ from  a material of interest.
    Fluctuations in the material generate fluctuating fields at the qubit's location which lead to its decoherence.
    {\bf b)} Using Ramsey spectroscopy or other generalized spin echo pulse sequences~\cite{vandersypen:2005,choi:2020a}, the qubit's decoherence dynamics can be used to  characterize the sample's fluctuations.
    {\bf c)}
    A summary of the frequency and length scales accessible to different experimental techniques highlights the complementarity of our proposed qubit-based noise spectroscopy ($T_2$ noise). 
    Techniques depicted include:
    Brillouin light scattering (BLS)~\cite{AdvancesBLS}, magneto-optical Kerr effect (MOKE)~\cite{haider2017reviewMOKE}, Raman spectroscopy~\cite{tan:2019}, inelastic neutron scattering (INS)~\cite{price2013Neutron}, nuclear magnetic resonance (NMR)~\cite{hore2015NMR,berthier2017NMR}, electron spin resonance~ \cite{katsumata2000ESR}, and muon spin resonance ($\mathrm{\mu SR}$)~\cite{hillier2022muon}. 
    {\bf d)} Schematic of the qubit's decoherence dynamics $C(\tau)=e^{-2\langle \phi^2\rangle}$: as the sample is tuned towards the critical point $\lambda = \lambda_c$ (darker colors, inset), enhanced fluctuations result in a shorter decoherence time $T_2^*$.
    More broadly, the decoherence rate and the shape of the profile encode characteristics of the phase transition, including its location, its critical exponents, and the presence of additional symmetries, Fig.~\ref{fig:DecaySignal}.
    }
    \label{fig:schematic} 
\end{figure}

In this Letter, we propose and analyze noise spectroscopy as a complementary probe of phase transitions and critical phenomena at low frequencies and momenta (Fig.~\ref{fig:schematic}c). 
Our central result is that such spectroscopy enables one to directly characterize the universality class of both classical and quantum phase transitions.
In particular, we introduce a simple method to quantitatively extract critical exponents via the scaling collapse of a probe qubit's decoherence profile (i.e.~as a function of experimentally tunable control parameters).

We highlight the flexibility and power of this approach in three distinct contexts.
First, we discuss how the decoherence profile captures the presence of symmetries and conservation laws.
Second, we show that the scaling behavior of the noise near the transition can efficiently characterize both thermal and quantum phase transitions. 
Finally, we demonstrate how our protocol is able to extract critical exponents---even when the probe qubit is not directly sensitive to the order parameter of the phase transition.

Let us begin by introducing the setup considered throughout this work (Fig.~\ref{fig:schematic}a):~a single isolated qubit (or a qubit ensemble) is located at a distance $d$ above a sample of interest.
    The qubit's energy splitting is given by $\Delta_0$ (with quantization axis $\hat{\n}$) and it couples to a time-dependent local field $\B(t)$ with strength $\gamma$: $\mbox{H}(t) = \frac{\Delta_0}{2} \hat{\n} \cdot \bm{\sigma}+ \frac{\gamma \B(t)}{2}\cdot \bm{\sigma}$,
where $\sigma^\alpha$ are Pauli operators \cite{fnUnits}.
While our discussions are applicable to generic qubit platforms (i.e.~solid-state spin defects, neutral atoms, trapped ions, superconducting qubits, etc.) coupled to a fluctuating field (i.e. magnetic, electric, strain, etc.), to be specific, we will describe our results in the context of a spin qubit coupled to a fluctuating magnetic field.
Three key ingredients relate the properties of the sample to the decoherence dynamics of the probe spin: 
(i) fluctuations within the material, 
(ii) the geometry, and
(iii) the measurement scheme (i.e.~pulse sequence).
The fluctuations of the field source $O^\alpha$ within the sample (e.g.~current density $j^\alpha$ or spin-density $s^\alpha$) can be directly characterized by the dynamic structure factor:
\beq    
S^{\alpha \beta}(\q,\omega) = \int dt\int d\r~ e^{i(\omega t - \bm{q}\cdot \bm{r})} \langle O^\alpha(t, \bm{r}) O^\beta(0,0)\rangle_T,~~~~
\label{eq:DSF}
\eeq
where $\omega$ is their frequency, $\q$ their momentum and 
$\langle \cdot \rangle_T$ corresponds to the thermal expectation at temperature $T$.

The relationship between the dynamics of these sources, $O^\alpha$, and the fluctuating magnetic field at the probe spin location, $\B(t)$, is determined by the system's geometry.
This role of geometry is best understood as a \emph{momentum filter function} $W^{\alpha \beta}_{d}(\hat{\mathbf{n}}, q)$ on the sample's fluctuations.
Because $W^{\alpha \beta}_{d}(\hat{\mathbf{n}}, q)$ is peaked around $q \sim 1/d$, the distance between the qubit and the sample provides direct control of the qubit's sensitivity to the momenta $\q$ of the fluctuations~\cite{supp}.
Simultaneously, different quantization axes $\hat{\mathbf{n}}$ allow one to extract different tensor components of the dynamic structure factor.
Akin to geometry, the measurement scheme induces a particular filter function, albeit in frequency space, $W_\tau(\omega)$ \cite{supp}.
Previous work has focused on $T_1$-based noise spectroscopy~\cite{Kolkowitz,Agarwal2017,Joaquin18,FT2018,CRD18,Rustagi,Du,CD2021,DC2021,AndersenDwyer,Sahay2021,Khoo,McLaughlin2022}; where the qubit is prepared along its quantization axis and its subsequent depolarization dynamics are determined by noise at frequency $\omega = \Delta_0$, leading to a sharp frequency filter function $W_\tau(\omega) \sim \delta(\omega-\Delta_0)$.
Given our focus on low frequency behavior, we turn to dephasing-based noise spectroscopy, best exemplified by Ramsey spectroscopy.
In this case, the qubit is prepared in a superposition $\ket{\psi} \propto \ket{\uparrow}+\ket{\downarrow}$ along the equator of the Bloch sphere, Fig.~\ref{fig:schematic}b.
In each experimental run of duration $\tau$, the magnetic field along the quantization axis, $\hat{\n}\cdot \B(t) = B(t)$, causes the qubit to Larmor precess by an angle $\phi = \gamma \int_0^\tau dt~B(t)$, (Fig.~\ref{fig:schematic}b)~\cite{fnOtherComp}.
Upon averaging over many experimental runs, the resulting density matrix exhibits an off-diagonal term given by $\langle e^{-i 2\phi} \rangle \approx e^{-2\langle\phi^2\rangle}$~(which precisely characterizes the qubit's decoherence).
If instead a $\pi$-pulse is applied in the middle of the Larmor precession, the phase accumulated becomes $\phi_{\mathrm{echo}} = \gamma \left[\int_0^{\tau/2} dt~B(t) - \int_{\tau/2}^{\tau} dt~B(t)\right]$, altering the qubit's sensitivity to different frequencies.
This effect is precisely captured by the frequency filter function, $W_\tau(\omega)$: for Ramsey spectroscopy, $W_\tau^{(\text{Ramsey})}(\omega) \propto \omega^{-2} \sin^2(\omega \tau/2)$ is peaked around $\omega=0$ with width $1/\tau$.
More intricate pulse sequences, such as spin echo and CPMG, can be used to tailor the properties of the filter function~\cite{supp,fnPulse,vandersypen:2005,choi:2020a}.

Bringing all these elements together, the qubit's decoherence profile, $e^{-2\langle\phi^2\rangle}$, depends on the pulse sequence [via $W_\tau(\omega)$], the geometry  [via $W^{\alpha \beta}_d(\hat{\mathbf{n}},q)$] and the properties of the sample of interest [via $S^{\alpha\beta}(q,\omega)$], and can be cast into a simple formula:
\begin{align}
\hspace{-3mm} 
\langle \phi^2\rangle = \int_{-\infty}^{\infty} \frac{d\omega}{2\pi}~W_{\tau}(\omega) \underbrace{\int_0^\infty \frac{dq}{2\pi} ~ W^{\alpha\beta}_d(\hat{\mathbf{n}}, q) S^{\alpha\beta}(q, \omega)}_{\mathcal{N}(\omega)}.
\label{eq:phiSq}
\end{align}
Here, $\N(\omega)$ denotes the noise spectral density of the fluctuating magnetic field $B(t)$ generated by the sample.

\begin{figure}[t]
    \centering
    \includegraphics[width = 3.3in]{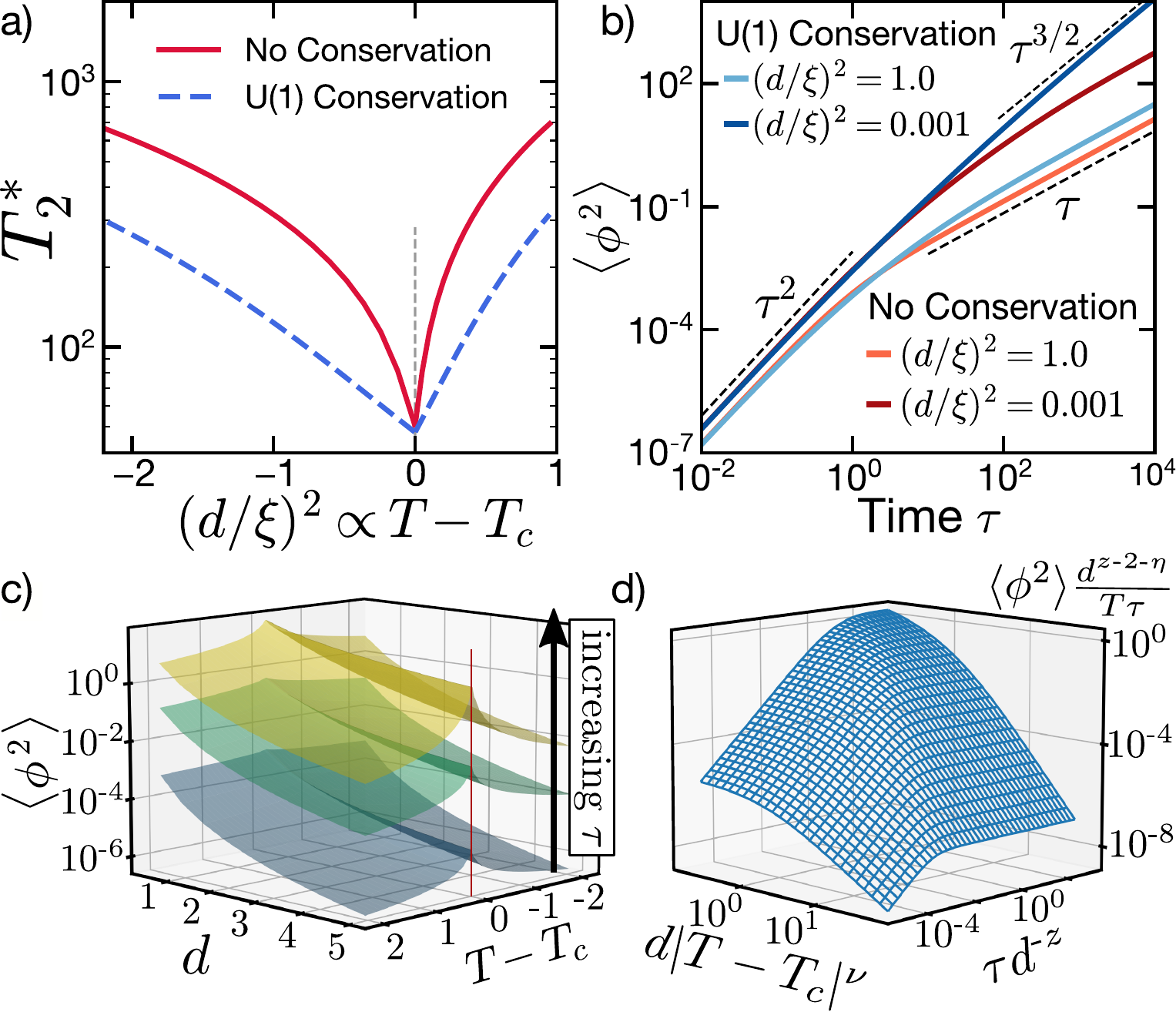}
    \caption{
    {\bf a)} Behavior of the Ramsey decoherence time, $T_2^*$, across a phase transition.
    $T_2^*$ exhibits a sharp feature at the transition, whose details are determined by the presence (or absence) of a conservation law in the order parameter.
    {\bf b)} Decoherence dynamics in the presence of different symmetries (with or without a conserved order parameter) and near or far from criticality [$(d/\xi)^{2}=0.001$ and $(d/\xi)^{2} = 1$]. 
    The presence of the conservation law  modifies the late-time  behavior of $\langle \phi^2 \rangle$---from $\tau\log\tau$ to $\tau^{3/2}$. 
    {\bf c)} Decoherence dynamics  as a function of the distance to sample $d$ and temperature $T$ for different times $\tau$.
    $\langle \phi^2\rangle$ is computed 
    for  Ramsey spectroscopy in the case of a classical Ising phase transition with no conservation law.
    {\bf d)} By performing a scaling collapse of $\langle \phi^2\rangle (d,\tau,T)$, one can directly extract  the critical exponents of the transition. 
    }
    \label{fig:DecaySignal}
\end{figure}    

\emph{Thermal phase transitions.---}Let us begin by exploring how noise spectroscopy enables the study of thermal (or classical) phase transitions. 
To be specific, we focus on Ramsey spectroscopy and spin models in two dimensions, where the momentum filter function takes the form $W_d(\hat{n},q) \sim  q^3 e^{-2 q d}$~\cite{supp}.
First, we consider the classical Ising transition in a two dimensional lattice of spins.
Although we are ultimately interested in the structure factor $S^{\alpha\beta}(\q,\omega)$ close to the transition, it is more convenient to analyze the behavior of the dynamical susceptibility $\chi^{\alpha\beta}(\q,\omega)$.
Fortunately, the two are intimately connected via the fluctuation-dissipation theorem:  $S^{\alpha\beta}(\q,\omega) = \frac{2T}{\omega} \text{Im}[\chi^{\alpha\beta}(\q,\omega)]$.
We focus on $\alpha=\beta=\mathrm{z}$ as it captures the critical correlations of the Ising order parameter --- the coarse-grained magnetization density.
To understand the fluctuations of the order parameter, we need to consider the slow relaxation dynamics toward equilibrium. 
These dynamics are dominated by long-wavelength fluctuations, and can be accounted for by a simple phenomenological model, $\chi^{-1}(\omega,\q) = \chi^{-1}(\q) - \frac{i\omega}{\Gamma(\q)}$ \cite{HH77,CardyBook,HH69,HHM1,HHM2}, where $\Gamma(\q)$ is the relaxation rate of the $\q$-Fourier mode.
As one approaches the critical temperature $T_\textrm{c}$, the order-parameter correlation length, $\xi$, diverges as $|T - T_\textrm{c}|^{-\nu}$; within mean-field theory $\nu = 1/2$ and $S^{\mathrm{zz}}(\q,\omega)$ becomes:
\beq
    \label{eq:S}
   S^{\mathrm{zz}}(\q,\omega) &=& \frac{2 T \Gamma(\q)}{\Gamma(\q)^2 J^2(\xi^{-2} + q^2)^2 + \omega^2}~.
\eeq
As perhaps expected, independent of the details of $\Gamma(\q)$, the magnetic field noise exhibits 
critical enhancement as one approaches the transition --- $S^{\mathrm{zz}}(\q, \omega)$ monotonically increases as the correlation length diverges and the minimum of the decoherence time $T_2^*$ diagnoses the critical temperature (Fig.~\ref{fig:DecaySignal}a). 

Crucially, noise spectroscopy enables the direct and quantitative characterization of critical phenomena, making it particularly amenable for studying transitions that deviate from the mean-field expectation.
More specifically, scaling considerations enable the derivation of an expression for $\langle \phi^2 \rangle$:
\beq
\label{eq:scaling} 
\langle \phi^2 \rangle = \frac{T \tau}{d^{2 + \eta - z}} F\left( \frac{\tau}{d^z},\frac{d}{\xi} \right), \quad \xi \propto |T - T_\mathrm{c}|^{-\nu}
\eeq
where $\eta$ is the anomalous scaling exponent, $z$ is the dynamical exponent and the scaling function $F$ depends on details of the dynamics.
By varying the distance $d$, the time $\tau$ and the temperature $T$ (Fig.~\ref{fig:DecaySignal}c), one can obtain a two-parameter scaling collapse for $\langle \phi^2 \rangle$ which immediately determines the critical exponents (Fig.~\ref{fig:DecaySignal}d). 

{\emph{Symmetries and conservation laws.}---}In addition to extracting critical exponents, certain qualitative features of the transition can be obtained from the scaling behavior of $\langle \phi^2\rangle$ with $\tau$.
As an example, we demonstrate how our approach naturally distinguishes between transitions with and without order-parameter conservation; for concreteness, let us return to the 2D classical Ising transition within mean-field theory.

We begin by focusing on the late-time decoherence dynamics, $\omega_0 \tau \gg 1$, where $\omega_0$ is the width of the noise spectral density $\N(\omega)$ (Eq.~\ref{eq:phiSq}) \cite{supp}. 
When the order parameter  (i.e. the $\q=0$ spin component) is not conserved, 
its decay rate $\Gamma(\q=0)$ will be a non-zero constant $\Gamma_0$, leading to a dynamical exponent $z = 2$~\cite{fnZ}. 
As shown in Fig.~\ref{fig:DecaySignal}b, this implies that $\langle \phi^2 \rangle $ exhibits an (almost) linear scaling with $\tau$ (see Table~\ref{tab:NoiseTab}).
In contrast, when the order parameter is conserved, its decay rate $\Gamma(\q=0)$ must vanish.
As a result, the small $\q$ behavior of $\Gamma(\q)$ is quadratic, $\Gamma(\q) \sim \sigma_s \q^2$, corresponding to diffusive spin-correlations away from criticality. 
However, at criticality, the correlation length, $\xi$, diverges and the spin-diffusion constant, $D_s = \sigma_s \xi^{-2}$, goes to zero, leading to a dynamical exponent $z = 4$ as per Eq.~\eqref{eq:S}.
Crucially, this results in a non-analytic behavior of the noise-spectral density $\N(\omega) \sim 1/\sqrt{\omega}$ at low-frequencies, manifesting itself as a different late-time scaling, $\langle \phi^2\rangle \sim \tau^{3/2}$ (Fig.~\ref{fig:DecaySignal}b and Table~\ref{tab:NoiseTab}).

A few remarks are in order. 
First, at short times, $\omega_0 \tau \ll 1$, the material dynamics are essentially ``frozen'' and the qubit is only sensitive to the static components of the noise, leading to $\langle \phi^2\rangle \sim \tau^2$.
A complementary perspective is that the Ramsey filter function is significantly broader than the noise spectral density and the  probe integrates the response across all frequencies.
The location of the crossover between this early-time behavior and the late-time dynamics (Fig.~\ref{fig:DecaySignal}b)
provides a direct measurement of the correlation time associated with the material's intrinsic dynamics~\cite{davis2021probing,dwyer:2021}.
Second, the scaling of the qubit's decoherence time with distance $d$ also provides insight into the approach to criticality.
In particular, within the critical regime, $d \ll \xi$, the qubit's decoherence will be  relatively insensitive to the sample-probe distance.
In the opposite limit, $d \gg \xi$,  the qubit's decoherence exhibits a significantly stronger \emph{power-law} dependence on $d$ (see Table~\ref{tab:NoiseTab}).


\begin{table*}
\setlength\extrarowheight{10pt}
\setlength{\tabcolsep}{7pt}
    \centering
    \begin{tabular}{ |c|c|c|c|c|c|c| }
    \hline
    \makecell{Nature of \\transition}& \makecell{Phase Transition}  & \makecell{Paradigmatic\\ model} & \makecell{Conserved\\Quantity}  & \makecell{Accessible \\ critical\\ exponents}  & \makecell{Noise $\langle \phi^2 \rangle$ at \\ criticality} & \makecell{Noise $\langle \phi^2 \rangle$ away \\ from criticality} \\
    \hline \hline
     \multirow{2}{*}{\vspace{-5mm}Thermal} &  \multirow{2}{*}{\vspace{-5mm}\makecell{Paramagnet \\to\\ Ferromagnet}}&\makecell{Ising} & --- & $\nu, \eta, z$ & $\dfrac{T \tau}{\Gamma_0} \ln \left( \dfrac{\tau \Gamma_0}{d^2} \right)$ & $\dfrac{T \tau \xi^4}{\Gamma_0 d^4}$   \\[2ex] \cline{3-7}
     &  &\makecell{XXZ}  & $S^\mathrm{z}$ & $ \nu, \eta, z$ & $\dfrac{T \tau^{3/2}}{\sqrt{\sigma_s}}$ & $\dfrac{T \tau \xi^4}{\sigma_s d^2}$ \\[2ex]
    \hline
    \hline
    \multirow{2}{*}{\vspace{-5mm}Quantum} &   \makecell{Para to Ferro} &\makecell{Ising} &  ---  & $\eta, \nu, z$ & $ T^{(2 + \eta )/z} \tau \ln\left( \dfrac{c}{d T^{1/z} } \right)$ & $ \dfrac{\tau}{d^4} e^{- a_2 \delta/T} $\\[2ex] \cline{2-7}
     &  \makecell{Para to AFM} &\makecell{Heisenberg}  & $\S$ & $z \nu$ & $\dfrac{T^3 \tau}{d^2} $ &  $\dfrac{\tau}{d^2} e^{- a_3 \delta/T} $\\[2ex]
    \hline
    \end{tabular}
    \caption{Critical and non-critical scaling of $\langle \phi^2 \rangle$ for paradigmatic models of classical and quantum phase transitions \cite{CardyBook} with temperature $T$, distance $d$ and time $\tau$, in the limit $\omega_0 \tau \gg 1$. The scaling behavior of $\langle \phi^2 \rangle$ for the thermal phase transitions is presented within mean-field theory where $\eta = 0$. For quantum transitions away from criticality ($T \ll \delta$), we do not include pre-factors of $T$ (since the noise is dominated by the exponential $e^{-\delta/T}$)~\cite{supp}.}
    \label{tab:NoiseTab}
\end{table*}

\emph{Quantum phase transitions.---}Let us now turn our attention to quantum phase transitions, where the ground state exhibits an abrupt, qualitative change upon tuning some parameter $\lambda$ (e.g. pressure, electron density, etc.).
The key distinction from the thermal case is the sensitivity of the qubit's decoherence to the spectral gap, $\Delta$, which goes to zero at the quantum critical point, $\lambda=\lambda_\textrm{c}$.
As a result, $\langle \phi^2\rangle$ still exhibits a simple scaling function, $\Psi$, albeit with an additional scaling parameter $\Delta/T$:
\begin{equation}
\label{eq:QPTscaling}
\langle \phi^2 \rangle = T^{\frac{2 + \eta - z}{z}} \Psi \left(\Delta \tau, \Delta d^{1/z}, \frac{\Delta}{T} \right),\;\Delta \propto |\lambda - \lambda_\textrm{c}|^{z\nu}.
\end{equation}%
In direct analogy to the thermal case, the critical exponents can be extracted via a scaling data collapse of the decoherence as a function of any three of the four parameters: (i) temperature $T$, (ii) tuning parameter $\lambda$, (iii) sample-probe distance $d$ and (iv) time $\tau$.

Despite their similarities, the quantum and thermal phase transitions are distinguishable by the \emph{temperature} scaling of $\langle \phi^2\rangle$. 
Away from criticality ($\Delta/T \gg 1$), the noise is controlled by thermally activated excitations on top of the ground state and is exponentially suppressed in $\Delta/T$ leading to $\langle \phi^2 \rangle \sim e^{-a\Delta/T}$ for $a>0$.
Near the critical point ($\Delta/T \to 0$), $\Psi$ approaches a constant and thus $\langle \phi^2 \rangle$ scales as a power-law in temperature.
By contrast, in classical phase transitions, the noise will always exhibit power-law correlations with temperature (see Table~\ref{tab:NoiseTab}).

The ability for noise spectroscopy to probe quantum phase transitions is extremely generic---the qubit does not need to couple to the order parameter, nor does the ordered phase need to be gapped.
We highlight this flexibility by studying the noise spectroscopy of a different transition, namely, the continuous symmetry breaking transition between a quantum paramagnet and a colinear N\'{e}el antiferromagnet in the two-dimensional Heisenberg model \cite{CHN1,CHN2,THC89,CSY94,SY92,ssbook}.
The order parameter for this transition is the staggered magnetization density which oscillates at the lattice scale; since this lengthscale is significantly smaller than the sample-probe distance $d$, the qubit is insensitive to fluctuations of the order parameter.
Rather, its decoherence is determined by the long-wavelength fluctuations of the spin density, which remain conserved owing to the Hamiltonian's SO$(3)$ spin-rotation symmetry. 
Consequently, the dynamic structure factor takes a simple diffusive form, $S^{\mathrm{zz}}(\q,\omega) = \frac{2 T \chi_u D_s \q^2}{\omega^2 + (D_s \q^2)^2} $ (analogous to Eq.~\ref{eq:S}) with diffusion constant $D_s$ and uniform static susceptibility $\chi_u$;  at late times, the decoherence dynamics take on a simple form $\langle \phi^2 \rangle \propto \frac{T \tau}{d^2} \frac{\chi_u}{D_s}$ that depends only on the ratio $\frac{\chi_u}{D_s}$.
The key ingredient that determines both $\chi_u$ and $D_s$ on either side of this quantum transition is an intrinsic energy scale $\delta$ --- in the  antiferromagnet $\delta = \rho_s$ is the spin stiffness associated with creating spatially non-uniform textures of the order parameter, while in the  paramagnet $\delta = \Delta$ is simply the spectral gap.
Specifically, in the vicinity of the transition, both $D_s$ and $\chi_u$ are universal functions of the ratio between the energy scale $\delta$ and temperature $T$, and thus the decoherence is controlled by $\delta/T$ \cite{SY92,CSY94,ssbook}. 

We are now in a position to understand how noise spectroscopy can characterize this transition.
First, let us focus on the regime away from criticality ($T \ll \delta$).
In this case, decoherence becomes suppressed with temperature owing to a divergent diffusion coefficient $D_s \sim e^{\delta/T}$  and a non-divergent susceptibility $\chi_u$ \cite{supp}.
By fixing $T$ and $\tau$ and varying $\lambda$, the combined critical exponents, $z\nu$, can be extracted from the stretched exponential decay of $\langle \phi^2\rangle \sim  (\tau/d^2) e^{- a_3 |\lambda-\lambda_\mathrm{c}|^{z\nu} /T}$.
On the other hand, in the quantum critical regime ($T \gg \delta$ and $\lambda \sim \lambda_\textrm{c}$), temperature is the only relevant energy-scale that determines both $\chi_u$ and $D_s$, and thus, the scaling of the decoherence is simply a power-law in temperature, $\langle \phi^2 \rangle \sim T^{3}$ \cite{supp}.
This provides a clear, quantitative signature that the system is in the critical regime.

A few remarks are in order. First, the detection of critical exponents in the absence of direct coupling between the qubit and the order parameter can also be applied to thermal phase transitions.
%
Second, given the sensitivity of the low-energy physics to underlying symmetries, the ability to carefully probe such dynamics  could enable the diagnosis of symmetry-breaking interactions.
%
%
Third, we have thus far, restricted our analysis to Gaussian noise, where two-point correlations are sufficient to fully describe the probe qubit's decoherence. 
The presence of higher-order moments in the noise distribution can affect the dynamics, especially near the  transition, where $\xi \gtrsim d$.
Interestingly, one can obtain the corresponding scaling forms and isolate these non-Gaussian contributions via pulse-sequence engineering \cite{supp,Viola16,AS11,Sung19}.
%
Finally, for some materials, the surface and bulk degrees of freedom can exhibit distinct critical phenomena, altering the probe qubit's decoherence dynamics~\cite{binder1983phase, cardy1996scaling, Satcher}. 
In this context, the sample-probe distance determines the relative contributions of surface and bulk fluctuations to the qubit's decoherence and can be used to isolate and characterize each transition separately.
However, to quantitatively characterize both critical phenomena, a more nuanced scaling theory is required.

\emph{Experimental blueprint}.---%
Our protocol can be applied to investigate a variety of critical phenomena, ranging from ferroelectric ordering~\cite{Sahay2021,Martin2016} and structural phase transitions~\cite{Satcher,muller1971static,bianco2017second,Kaneko2013}, to magnetic ordering~\cite{Elliott_Magnetism, brando:2016, ghosh:2021} and superconductivity~\cite{yip:2019, lillie:2020,CD2021,DC2021}.
In what follows, we highlight our approach in the context of magnetic insulators probed via solid-state, electronic spin defects.
We focus on two particular defects --- the negatively charged boron vacancy ($V_\textrm{B}^{-}$) in hexagonal boron nitride (hBN)~\cite{kianinia:2020, gottscholl:2020, healey:2021, huang:2022a} and the Nitrogen-Vacancy (NV) color center in diamond~\cite{nvreview, schirhagl:2014a, casola:2018}  --- with the goal of highlighting their complementary operating modalities. 
The former can be created within flakes of hBN that are directly placed on a sample of interest~\cite{healey:2021}, while the latter can be embedded within the tip of a (diamond) cantilever and used as a scanning probe~\cite{nvsinglespin,pelliccione:2016,thiel:2019,vool:2021}.
Both defects feature spin $S=1$ electronic ground states with zero-field splittings in the $\sim$GHz regime (Fig.~\ref{fig:schematic}c). 
Two spin states (forming our probe qubit) can be isolated by either resolving the hyperfine interaction or via an  external magnetic field.
The frequency range of $T_2$-based noise spectroscopy, $\sim$kHz-MHz, is limited by the defect's local environment and the achievable Rabi frequency for pulsed control~\cite{Maze2011,Doherty2013,hong2013nanoscale,Mittiga2018, bauch:2020a}.

Both NV centers and $V_\textrm{B}^{-}$ defects are particularly well-placed for studying 
 two-dimensional magnetic materials.
 As an example, consider monolayer CrI$_3$, a two-dimensional magnet with an Ising transition at $T_\textrm{c}\approx 45~\mathrm{K}$ \cite{mcguire2015,gong2017,huang2017,Lado2017,liu2019}.
 We estimate that in the diffusive regime, the sampled-induced decoherence time will be approximately, $T_2^{\mathrm{echo}} \approx 5 ~\mathrm{\mu s}$ (using $T \approx 60~\mathrm{K}$ and $d=10~\mathrm{nm}$)~\cite{supp}, with critical fluctuations near the transition  further reducing this value \cite{fn1}.
Crucially, this additional noise  dominates over the NV's intrinsic decoherence, $T_2^{\mathrm{echo}} \sim 100~\mathrm{\mu s}$ \cite{Sangtawesin2019, bauch:2020a}, enabling its detection. 
Zooming out, the broader landscape of van der Waals heterostructures and moir\'{e} materials
offers a wide range of correlated insulators and magnetic transitions to explore~\cite{andrei2021marvels,mak2022semiconductorMoire}.
For example, pressure-driven quantum critical points in two dimensional materials, such as FeSe~\cite{KreiselFeSe,lee2018routesFeSe}, as well as strain-tuned magnetic transitions in monolayer metallic halides, such as the aforementioned CrI$_3$~\cite{wu2019strain}, can be accessed by directly incorporating the NV center into diamond anvil cells \cite{Satcher,lesik2019,yip:2019}, while magnetic domain formation can be imaged \emph{in situ} by directly incorporating $V_\textrm{B}^{-}$  into the hBN that encapsulates many 2D materials~\cite{healey2021,huang2022}. 
Simultaneously, the detection and characterization of spin and charge fluctuations can elucidate the nature of the magnetic order in twisted bilayer graphene \cite{AndreiMacdonaldReview,Sharpe2019,Serlin2020,BCZ2020,Zhang2019,NickPRX,LianTBG4}, as well as shed light into the continuous Mott transition recently observed in moir\'{e} transition metal dichalcogenides \cite{Li2021,Musser2021,mak2022semiconductorMoire}. 

\emph{Acknowledgements}.---We gratefully acknowledge discussions with and the insights of K. Akkaravarawong, E. Davis, Z. Dai, S. Hsieh, J. R. Nieva, S. Whitsitt, M. P. Zaletel and C. Zu. 
This work was supported in part by the U.S. DOE, Office of Science, Office of Advanced Scientific Computing Research, under the Accelerated Research in Quantum Computing (ARQC) program, the
US Department of Energy (Award No.~DE-SC0019241), the ARO through the MURI program (grant number W911NF-17-1-0323), the David and Lucile Packard Foundation, the Alfred P.~Sloan foundation.
F.M. acknowledges support from the NSF through a grant for ITAMP at Harvard University. 
E.D. acknowledges support from the Swiss National Science Foundation under Division II.
\bibliography{references}

\begin{thebibliography}{115}
\bibitem [{fnU()}]{fnUnits}%
  \BibitemOpen
  \href@noop {} {}\bibinfo {note} {Throughout this work we use the natural unit
  convention where $k_b = \hbar =1$.}\BibitemShut {Stop}%
\bibitem [{fnO()}]{fnOtherComp}%
  \BibitemOpen
  \href@noop {} {}\bibinfo {note} {The other components of the magnetic field
  induce depolarization on the spin, which can be accounted independently by
  the aforementioned $T_1$ spectroscopy \cite{supp}.}\BibitemShut {Stop}%
\bibitem [{fnP()}]{fnPulse}%
  \BibitemOpen
  \href@noop {} {}\bibinfo {note} {Note that for generalized echo sequences,
  i.e.~CPMG~\cite{carr:1954}, the total duration, $\tau$, is given by the
  number of pulses, $N_\textrm{p}$, multiplied by the interpulse spacing,
  $\tau_\textrm{p}$~\cite{supp}.}\BibitemShut {Stop}%
\end{thebibliography}

\end{document}


\title{Supplementary Material: Quantum noise spectroscopy of dynamical critical phenomena}

\author{Francisco Machado}
\affiliation{ITAMP, Harvard-Smithsonian Center for Astrophysics, Cambridge, Massachusetts, 02138, USA}
\affiliation{Department of Physics, Harvard University, MA 02138, USA}
\affiliation{Department of Physics, University of California, Berkeley, CA 94720, USA}

\author{Eugene A. Demler}
\affiliation{Institute for Theoretical Physics, ETH Zurich, 8093 Zurich, Switzerland}

\author{Norman Y. Yao}
\affiliation{Department of Physics, Harvard University, MA 02138, USA}
\affiliation{Department of Physics, University of California, Berkeley, CA 94720, USA}

\author{Shubhayu Chatterjee}
\affiliation{Department of Physics, University of California, Berkeley, CA 94720, USA}
\affiliation{Department of Physics, Carnegie Mellon University, Pittsburgh, PA 15213, USA}

\maketitle

\section{Defining the filter functions}
In this section, we define the frequency and momentum filter functions [$W_\tau(\omega)$ and $W^{\alpha \beta}_d(\hat{\n},q)$ respectively], and derive the central equation for decoherence dynamics of the spin qubit --- Eq.~(2) in the main text.
 
We start with the qubit Hamiltonian:
\begin{equation}
\mbox{H}(t) = \frac{\Delta_0}{2} \hat{\n} \cdot \bm{\sigma}+ \frac{\gamma \B(t)}{2}\cdot \bm{\sigma}
\end{equation} 
and start by considering a simple Ramsey sequence. 
In this case, the qubit is initialized at $t=0$ in the superposition $\ket{+} = (\ket{\uparrow} + \ket{\downarrow})/\sqrt{2}$, and then allowed to evolve under a time dependent dephasing field $B(t) = \bm{B}(t) \cdot \hat{\bm{n}}$ that points along the quantization axis $\hat{\bm{n}}$---we assume that the zero-field splitting $\Delta_0$ is much larger than the local field $\gamma \bm{B}(t)$. 
In the rotated frame of reference (w.r.t. the zero-field splitting $\Delta_0$ of the qubit), the density matrix at time $t$ is given by:
\beq
\rho(t) = \frac{1}{2}\left( e^{i\phi} \ket{\uparrow} + e^{-i\phi} \ket{\downarrow} \right)\left( e^{-i\phi} \bra{\uparrow} + e^{i\phi} \bra{\downarrow} \right), \text{ where } \phi(t) = \frac{\gamma}{2\hbar} \int_0^{t} dt^\prime \, B(t^\prime) 
\label{eq:rhot}
\eeq
At $t = \tau$, the off-diagonal (coherence) element of the density matrix is measured, and this measurement is repeated for many realizations of the magnetic field noise.  
From Eq.~\eqref{eq:rhot}, this is given by $\langle e^{2i \phi (\tau)} \rangle$.
Assuming that the probability distribution for magnetic field fluctuations is Gaussian with zero-mean, we can write the ensemble average as:
\beq
\langle e^{2i \phi (\tau)} \rangle = e^{- 2 \langle \phi^2(\tau) \rangle}, \text{ where } \langle \phi^2(\tau) \rangle = \left( \frac{\gamma}{2\hbar} \right)^2 \int_0^{\tau} dt_1 \, \int_0^{\tau} dt_2 \, \langle B(t_1) B(t_2) \rangle
\eeq

\subsection{Frequency filter function}
If the magnetic field fluctuations arise from spin fluctuations in a nearby sample in thermal equilibrium, we expect the magnetic field correlations to be time-translation invariant.
This implies that the correlation function can be written in terms of a noise spectral density $\mathcal{N}(\omega)$ as follows:
\beq
\langle B(t_1) B(t_2) \rangle = \int_{-\infty}^{\infty} \frac{d\omega}{2\pi} \, e^{-i\omega(t_1 - t_2)} \mathcal{N}(\omega)
\label{eq:Bcorr}
\eeq
Using this relationship, we can relate $\langle \phi^2 \rangle$ (suppressing the argument $\tau$ for ease of notation) to the noise spectral density as:
\begin{align}
&\langle \phi^2 \rangle = \int_{-\infty}^{\infty} \frac{d\omega}{2\pi} \, W_\tau(\omega)\mathcal{N}(\omega) \\
\text{ where for Ramsey/}T_2^*: & \quad W_\tau(\omega) = \left( \frac{\gamma}{2\hbar} \right)^2  \bigg| \int_0^{\tau} dt \, e^{i \omega t} \bigg|^2 = \left( \frac{\gamma}{2\hbar} \right)^2 \left( \frac{4\sin^2(\omega \tau/2)}{\omega^2} \right)  ~~~
\end{align}
Note that for $T_2^*$, the frequency filter function $W_\tau(\omega)$ is peaked at $\omega = 0$, with peak height $\sim \tau^2$ and width $\sim 1/\tau$. 
Thus, the decay of coherence, characterized by $\langle \phi^2 \rangle$ is mainly determined by the spectral density of $\mathcal{N}(\omega)$ for $|\omega| \lesssim 1/\tau$, as argued in the main text. 

\begin{figure}
    \centering
    \includegraphics[width = \textwidth]{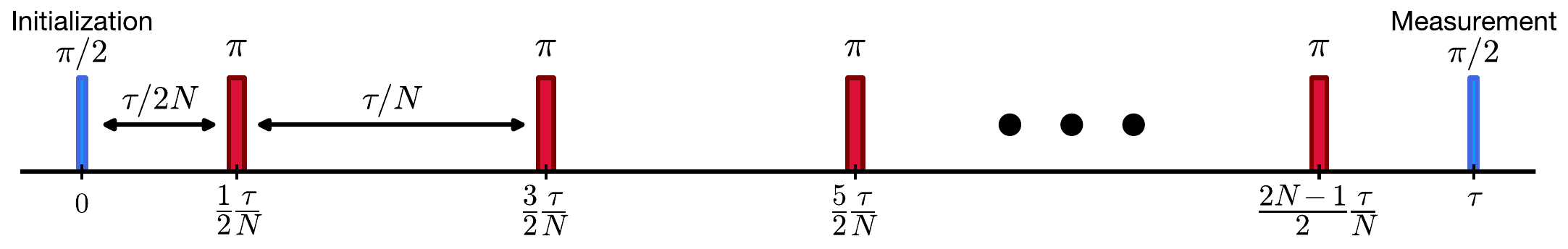}
    \caption{Schematic of the CPMG pulse sequence. After initialization, the first $\pi$-pulse is applied after $\tau/2N$ time has elapsed. Afterwards, a $\pi$ pulse is applied every $\tau/N$ until the measurement is performed at $\tau$. }
    \label{fig:CPMG}
\end{figure}

We now proceed to derive the filter function for more complicated pulse sequences, which enable sharper frequency-selective of $\mathcal{N}(\omega)$. 
For simplicity, we will neglect the possibility of imperfection in these pulses, i.e, we will assume them to be instantaneous and exact. 
While such imperfections are necessarily present in any experiment, the details of the pulse sequences can be chosen to minimize the effect of imperfections \cite{choiRobustDynamicHamiltonian2020, zhouQuantumMetrologyStrongly2020}, and this allows us to focus on a perfectly executed pulse sequence to derive simple analytic results. 
Specifically, let us consider the CPMG sequence --- a generalized spin-echo sequence of $N$ instantaneous $\pi$-pulses about the $\hat{x}$ axis (initial direction of the qubit spin), separated from each other by $\tau/N$ and separated from the start and end of the sequence by $\tau/(2N)$ (see Fig.~\ref{fig:CPMG}).
Each such $\pi$-pulse, in the rotating frame of the qubit, changes the sign of the Hamiltonian ($H \to -H$), so that the phase accumulation is \emph{negative} after an odd number of $\pi$-pulses and positive otherwise.
Accordingly, the sequence filter function is given by:
\begin{align}
    W_\tau(\omega) &=  \left(\frac{\gamma}{2\hbar}\right)^2\left| \left(\int_0^{\tau/{2N}} dt~ e^{-i\omega t}\right) + \left(\sum_{n=1}^{N-1} (-1)^n \int_{ \frac{\tau}{N}(n-1/2)}^{\frac{\tau}{N}(n+1/2)} dt~ e^{-i\omega t} \right)+ (-1)^N \left(\int_{\tau-\tau/{2N}}^\tau dt~ e^{-i\omega t}\right)\right|^2 = \\[10pt]
    &= \left(\frac{\gamma}{2\hbar}\right)^2 \dfrac{16}{\omega^2}\dfrac{ \sin^4\frac{\omega\tau}{4N}}{\cos^2 \frac{\omega\tau}{2N}}\begin{cases} 
    \cos^2 \frac{\omega\tau}{2} & N~\textrm{odd}  \\[6pt]
    \sin^2 \frac{\omega\tau}{2}  & N~\textrm{even} \\
    \end{cases}
\label{eq:FreqFilter}
\end{align}
The $N = 1$ case corresponds to Hahn-echo ($T_2$), with $W_\tau(\omega) = (2\gamma/\hbar)^2 \sin^4(\omega \tau/4)/\omega^2$, which is peaked at $\omega \sim \pi/\tau$ with peak height $\sim \tau^2$ and width $\sim 1/\tau$.
We can also consider the large $N$-limit at fixed interpulse frequency $\omega_p = \pi N /\tau$. In this regime, $W_\tau(\omega)$ is sharply peaked at $\omega_p$, with width $\sim \omega_p/N$.
In this limit, we can approximate $W_\tau(\omega)$ by a sum of Dirac-delta functions as follows:
\beq
\lim_{\substack{N,\tau \to \infty\\N\pi/\tau = \omega_p}} W_{\tau}(\omega) \; = \; \tau \left( \frac{\gamma}{\hbar} \right)^2 \left( \frac{2}{\pi} \right) \sum_{n=0}^\infty \frac{ \delta\left[\omega - (2n+1)\omega_p \right]}{(2n+1)^2} 
\label{eq:largeNFreqFilter}
\eeq
where we have fixed the overall coefficient by demanding that our approximation conserves the total weight $\int d\omega \, W_\tau(\omega)$.
This form of the frequency filter function substantiates the claim of enhanced frequency selectivity via CPMG sequences in the main text.
The presence of higher harmonics stems from the simplicity of the CPMG pulse sequence; by studying more complicated pulse sequences one can further isolate the lower frequency $\delta$-function from the higher harmonics.
Alternatively, one can compare the decoherence dynamics under different pulse sequences and isolate the low frequency contribution~\cite{Sung19}.

Let us conclude by writing down the general expression of the filter function. In general, each application of a $\pi$-pulse changes the sign of the Hamiltonian in the qubit's rotating frame. Let us define $f(t)$ as the function that keeps track of the sign of the Hamiltonian; at time $t$, $f(t) = +1$ if the sign of the Hamiltonian is positive and $f(t) = -1$ if the sign of the Hamiltonian is negative.
Then the filter function is then simply given by:
\begin{equation}
    W_\tau(\omega) = \left(\frac{\gamma}{2\hbar}\right)^2 \left | \int_0^\tau dt~ e^{-i\omega t} f(t) \right|^2
\end{equation}

\subsection{Momentum filter function}
Next, we derive the momentum filter function $W^{\alpha \beta}_d(\hat{\n},q)$, which characterizes relates the fluctuations in the sample $S^{\alpha\beta}(\bm{q},\omega)$ to the noise spectral density $\mathcal{N}(\omega)$ of the local field $B(t)$.
Therefore we need an appropriate quantum generalization of the classical correlations of the magnetic field in Eq.~\eqref{eq:Bcorr}, which may be obtained by replacing the classical correlator by the quantum anticommutator as follows:
\beq
\mathcal{N}(\omega) = \int_{-\infty}^{\infty} dt\, e^{i \omega t}\langle B(t) B(0) \rangle \to \frac{1}{2} \int_{-\infty}^{\infty} dt\, e^{i \omega t} \langle \{ \hat{B}(t) , \hat{B}(0) \} \rangle
\label{eq:Nquantum}
\eeq
In Eq.~\eqref{eq:Nquantum}, we denoted the field by $\hat{B}$ to emphasize that these are operators that relate to fluctuations of the quantum sources in the sample, that may behave classically or quantum mechanically.
However, in the rest of the text we will simply write $B$ and the classical/quantum limit is to be understood from the context. 
For computational convenience, we use the relation between the noise spectral density to the retarded correlation function via the fluctuation-dissipation theorem for temperature $T$ ($\beta = 1/k_B T$). 
\beq
\mathcal{N}(\omega) = \hbar \, \coth\left( \frac{\beta \hbar \omega}{2} \right) \left( - \text{Im}\left[ C_{BB}(\omega) \right] \right)
\eeq
where $C_{BB}(\omega) = \int_{-\infty}^{\infty} dt\, e^{i \omega t} C_{BB}(t)$ is the Fourier transform of the retarded field-field correlation function $C_{BB}(t) = - i \theta(t) \langle [B(t), B(0)] \rangle_{\beta}$. What remains is to relate the local field $B(t)$ correlations to the source correlations in the sample.
To this end, we specialize to the case where the local field is magnetic in nature and arises from fluctuation of spins or currents in the sample---while different fields will exhibit different propagators, this does not alter the prescription presented here.
In this case, the fields and the sources are related via Maxwell's equations; we turn to their analysis next. 

Both electrical current-fluctuations and spin-fluctuations in the sample can act as a source for magnetic field noise. 
For simplicity, we focus on insulating materials with a large gap to charged excitations in the temperature range of interest, and consider the dipolar magnetic field generated by spin-degrees of freedom $\S(\r_i)$ on a lattice (we scale each spin by $\hbar$ so that $\S(\r_i)$ is dimensionless). 
\beq
\B_\alpha(\r_{\text{qubit}}, t) = \mu_0 \mu_B g_s \sum_{\r_i} H_{\alpha \beta}(\r_{\text{qubit}} - \r_i) S^\beta(\r_i,t), \text{ where } H_{\alpha \beta}(\r) = \frac{1}{4\pi} \left( \frac{3 r_\alpha r_\beta - r^2 \delta_{\alpha \beta} }{r^5} \right)
\eeq
Note that we have neglected retardation effects, i.e, taken the limit of infinite speed of light $c$ so that the propagator $H_{\alpha \beta}$ is instantaneous (for further discussion on this point, see Ref.~\onlinecite{CRD18}). 
Assuming a large two-dimensional sample at the x-y plane, we can take advantage of translation invariance: this allows us to position the qubit at $\r_{\text{qubit}} = (0,0,d)$ and move to momentum space to calculate correlations. 
To this end, we define spatial Fourier transforms for two dimensional momentum $\q$, as follows:
\beq
B_\alpha(\r_{\text{qubit}}, t) &=& \frac{1}{\sqrt{N}} \sum_{\q} e^{-i \q \cdot \r_{\text{qubit}}} B_\alpha(\q,t), ~~~~~~ H_{\alpha \beta}(\r_{\text{qubit}} - \r_i) = \frac{1}{N} \sum_{\q} e^{-i \q \cdot (\r_{\text{qubit}} - \r_i )} H_{\alpha \beta}(\q), \text{ and} \nn S_\beta(\r_i, t) &=& \frac{1}{\sqrt{N}} \sum_{\q} e^{-i \q \cdot \r_i} S_\beta(\q,t), ~~~ \text{ such that } B_{\alpha}(\q,t) = \mu_0 \mu_B g_s H_{\alpha \beta}(\q) S^\beta(\q,t)
\eeq
where $N$ decribes the number of sources in the system and $H_{\alpha \beta}(\q)$ may be found by Fourier transforming by the dipolar propagator (see Ref.~\onlinecite{Joaquin18} for details):
\beq
H_{\alpha \beta}(\q) = \frac{e^{- |\q| d}}{2 a^2} \begin{pmatrix} \frac{q_x^2}{|\q|} & \frac{q_x q_y}{|\q|} & i q_x \\ \frac{q_x q_y}{|\q|} & \frac{q_y^2}{|\q|} & i q_y \\ i q_x & i q_y & - |\q| \end{pmatrix}
\label{eq:MaxwellKernel}
\eeq
where $a$ is the lattice spacing (assuming square lattice). 
For concreteness we set the qubit quantization axis $\hat{\n} = \hat{\mathrm{z}}$, which implies that the relevant field correlator take the following form:
\beq
\langle \{ B(t) , B(0) \} \rangle = \frac{1}{N} \sum_{\q} H_{\mathrm{z} \alpha}(\q) H_{\mathrm{z} \beta}(-\q) \langle \{ S^\alpha(\q, t) S^\beta(-\q,0) \} \rangle_\beta
\eeq
where we have used translation invariance of the spin-spin correlation functions. 
Further assuming rotational invariance of low-energy correlations (or conservation of total $S^\mathrm{z}$ by the sample Hamiltonian, see Ref.~\onlinecite{CRD18}), we find the following the noise spectral density in the continuum limit only depends on three terms. 
\beq
\mathcal{N}(\omega) = \frac{\hbar (\mu_0 \mu_B g_s)^2}{4 a^2} \coth\left( \frac{\beta \hbar \omega}{2} \right)  \int_0^{\infty} \frac{dq}{2\pi} \, q^3 e^{-2qd} \left( \text{Im} \left[ \frac{1}{2}( \chi_{+-}(q, \omega) + \chi_{-+}(q, \omega) ) + \chi_{\mathrm{zz}}(q, \omega) \right] \right)
\label{eq:N}
\eeq
where we have used the conventional definition of the retarded spin-spin correlation function \cite{coleman_2015}:
\beq
\chi_{\alpha \beta}(\q, \omega + i 0^+) = \int_{-\infty}^{\infty} dt\, e^{i (\omega + i 0^+) t} \chi_{\alpha \beta}(\q, t), \quad \text{ and }  \quad \chi_{\alpha \beta}(\q, t) =  \frac{i \theta(t)}{\hbar} \langle [S^\alpha(\q,t), S^\beta(-\q,0)] \rangle_{\beta}
\eeq
We will mostly be interested in critical fluctuations of $S^\mathrm{z}$, and therefore often use the shorthand $\chi_{\mathrm{zz}}(\q, \omega + i0^+) = \chi(\q,\omega)$ and neglect the $\chi_{+-}/\chi_{-+}$ terms in Eq.~\eqref{eq:N}. 
These latter terms are known to be important for detecting quasiparticle physics away from the critical point \cite{CRD18,Joaquin18}, but are expected to be subdominant to $\chi_{zz}$ in the vicinity of the critical point.
Further, in the experimentally relevant regime, we always have $\beta \hbar \omega \lesssim 10^{-2}$, since the largest frequencies considered, $\omega \lesssim 1$~GHz, are much lower than the smallest temperatures, $T \sim 1$~K.
In this limit, the retarded spin-spin correlator and the dynamic spin-structure factor $S(\q,\omega) \equiv S_{\mathrm{zz}}(\q, \omega)$ are simply related by the fluctuation dissipation theorem \cite{coleman_2015,ssbook}:
\beq
S(\q,\omega) =  \int_{-\infty}^{\infty} dt \sum_{\r_i} e^{i (\omega t - \q \cdot \r_i)} \langle S_\mathrm{z}(\r_i,t) S_\mathrm{z}(\bm{0},0) \rangle = \left( \frac{2\hbar}{1 - e^{-\beta \hbar \omega}} \right) \left(  \text{Im}[ \chi(\q,\omega)] \right) \xrightarrow{\beta \hbar \omega \to 0} \frac{2}{\beta \omega}\left(   \text{Im}[ \chi(\q,\omega)] \right) 
\label{eq:Sqw}
\eeq
Using these approximations further allows us to relate the noise spectral density $\mathcal{N}(\omega)$ to the dynamic spin-structure factor $S(q,\omega)$ via a momentum filter function $W_{d}(q)$:
\beq
\mathcal{N}(\omega) = \int_0^\infty \frac{dq}{2\pi} \, W_d(q) S(q,\omega), \text{ where } W_d(q) \equiv W^{\mathrm{zz}}_d(\hat{\mathrm{z}},q) = \frac{(\mu_0 \mu_B g_s)^2}{4 a^2} q^3 e^{-2 q d}
\label{eq:Wdq}
\eeq
This completes our derivation of Eq.~(2) in the main text. 

For calculating the dynamic spin-structure factor, we will often use a continuum field-theory where we define $S(\q,\omega)$ with an integral over two-dimensional space $(\int d^2r)$ rather than a sum of lattice sites ($\sum_{\r_i}$), as in Eq.~\eqref{eq:Sqw}.
We can account for this simply by taking $W_d(q) \to W_d(q)/a^2$ in Eq.~\eqref{eq:Wdq} (since $\sum_{\r_i} \to \int d\r/a^2$ in continuum).
Finally, we comment that for a thin film with many  weakly-intercorrelated layers, the total noise can be found by simply summing the noise from each layer. 
Operationally, this amounts to taking $e^{-2qd} \to \sum_{\ell} e^{- 2 q d_\ell}$ in $W_d(q)$, where $d_\ell$ is the distance of the $\ell^{th}$ layer from the probe qubit.

\section{Effect of noise-induced $T_1$ depolarization}

Although our analysis has focused on the the decoherence dynamics of the probe qubit, the presence of noise at the resonance frequency leads to depolarization that affects the overall signal---this depolarization effect lies at the heart of previously proposed $T_1$ noise spectroscopy techniques~\cite{Kolkowitz,Agarwal2017,Joaquin18,FT2018,CRD18,Rustagi,Du,CD2021,DC2021,AndersenDwyer,Sahay2021,Khoo}. 

Crucially, such depolarization effect simply induces an overall decay of the observed signal, rather than imparting any additional features.
This relies on the fact that the depolarization decay depends on the bare frequency splitting of the qubit probe and not on the frequency ($\sim1/\tau)$) of the pulse sequence applied to study the decoherence dynamics.

To make this claim more precise, we can analyze the qubit probe dynamics using a master equation formalism that includes depolarization via a simple Linbladian term. We work in the rotating frame of the zero field splitting, whose eigenstates are $\{ \ket{g},\ket{e} \}$ (the ground and excited states, respectively); $\sigma_\mathrm{z}$ is Pauli operator that acts diagonally in this basis.
\begin{equation}
    \frac{d \rho}{dt} = - \frac{\gamma B(t)}{i2\hbar} [\sigma_\mathrm{z}, \rho] - \frac{1}{2T_1}\left\{ \ket{g}\bra{g}\rho+ \rho\ket{g}\bra{g} - 2 \ket{g}\bra{e}\rho\ket{e}\bra{g} + \ket{e}\bra{e}\rho+ \rho\ket{e}\bra{e} - 2 \ket{e}\bra{g}\rho\ket{g}\bra{e}\right\}
\end{equation}
Focusing on the dynamics of the coherence term $\rho_{12}$; when there is an equal population in both levels $\rho_{gg}=\rho_{ee} = \frac{1}{2}$ (as considered in our setup), we have:
\begin{equation} \label{eq:decwithT1}
    \frac{d\rho_{ge}}{dt} = \left[ - \frac{1}{T_1}  - i\frac{\gamma B(t)}{\hbar}\right] \rho_{ge}
\end{equation}

From Eq.~\eqref{eq:decwithT1}, it is straightforward to understand the effect of depolarization noise: on top of the dynamics generated by the local field $B(t)$, the coherence (measured by $\rho_{ge}$), experiences an overall decay rate $1/T_1$.
Since this depolarization rate can be independently measured (by preparing the system in either $\left|\uparrow\right\rangle$ or  $\left|\downarrow\right\rangle$ and watching the subsequent equilibration), the decoherence dynamics can be isolated.
In practice, the depolarization time, $T_1$, will induce an effective upper limit in the amount of time the decoherence dynamics can be investigated; beyond $T_1$, the signal is exponentially suppressed requiring an exponentially large number of measurements to extract it from the noise background.
 
\section{Noise at thermal transitions}
In this section, we elaborate on our discussion of decoherence in the vicinity of thermal transitions. We first provide a complimentary picture based on a real-space and real-time approach that allows a heuristic derivation of the scaling of Gaussian decoherence close to the transition for a Ramsey pulse sequence. We then explicitly derive the scaling functions within mean-field theory and generalize to include fluctuations beyond mean-field. Finally, we discuss generalizations to more complicated pulse sequences, using CPMG as a specific example. For all subsequent discussion, we have set $\hbar = 1 = k_B$.
\newline

\emph{Real-time, real-space picture:} Let us focus on thermal phase transitions in a two dimensional magnetic insulator a distance $d$ from the probe; in this case, noise arises from the dipolar fields of fluctuating spins. For a Ramsey ($T_2^*$) sequence, we then have:
\beq
\langle \phi^2 \rangle \sim \int_0^{\tau} dt \int_0^{\tau} dt^\prime \int_{\r \in \mathcal{A}} d\r \int_{\rp \in \mathcal{A}} d\rp \bigg\langle \frac{ S_\r(t)}{d^3} \frac{S_\rp( t^\prime)}{d^3} \bigg \rangle  =  \frac{\tau}{d^4}\int_0^{\tau} dt  \int d\r \, \langle S_\r(t) S_0(0) \rangle
\eeq
where we have used translation[time-translation] invariance to pull out a factor of $d^2$[$\tau$], and $\mathcal{A} \approx d^2$ indicates that the qubit placed at distance $d$ is (roughly) sensitive to an area of the order of $d^2$ on the sample. 
The spatial integral can then scale in two distinct ways: 
i) if the sample probe distance $d$ is larger than the spin-correlation length $\xi$ (typically the case away from the critical regime), then the spatial integral roughly contributes $\xi^2$, 
ii) if instead $d\ll \xi$, then it contributes $d^2$. 
Analogously, the temporal-integral can also exhibit two distinct scaling regimes:
i) if the time-scale $\tau$ of measurement is much smaller than the inverse typical frequency scale that sets the approach to equilibrium from small perturbations, $\tau \omega_0 \ll 1$, then we are insensitive to the dynamics and the qubit effectively measures equal-time correlations, $\langle S_\r(0) S_0(0) \rangle$;
ii) if instead, $\tau \omega_0 \gg 1$, the time-integral contributes $\int_0^\infty dt e^{- \omega_0 t} \sim \omega_0^{-1}$.
These considerations can be combined into a simple scaling for $\langle \phi^2 \rangle$ (when using the Ramsey sequence):
\beq
\langle \phi^2 \rangle \propto \frac{\tau^2}{d^2} \times \text{min}\left(1,\frac{1}{\omega_0 \tau} \right) \times \text{min}\left(1, \frac{\xi^2}{d^2}\right)
\label{eq:AppScR}
\eeq
While the above derivation assumed exponential decay of correlations, the long-time scaling in Eq.~(\ref{eq:AppScR}) may be modified when the correlations only fall-off as a power law $\sim (\omega_0 t)^{- \alpha}$ at large times with $0 < \alpha < 1$, implying that the time-integral contributes $\tau (\omega_0 \tau)^{-\alpha}$. For example, at mean-field level, the critical Ising model with $S^z$ conservation is characterized by $\alpha = 1/2$ which corresponds to $\langle \phi^2 \rangle$ scaling as $\tau^{3/2}/\sqrt{\omega_0}$ at large $\tau$.

Critical physics is characterized by a diverging correlation length $\xi \sim |T - T_\mathrm{c}|^{-\nu}$, and a vanishing energy scale $\Omega_0 \sim |T - T_\mathrm{c}|^{z \nu}$ corresponding to `critical slowing down'. The qubit is sensitive to a noise-spectral density which is usually a Lorentzian with width $\omega_0 \approx$ $\Omega_0$ in the large $d$ limit (with corrections of order $d^{-1}$ or higher).  Therefore, Eq.~(\ref{eq:AppScR}) implies that as we tune $T$ towards $T_\mathrm{c}$, $\omega_0$ decreases; correspondingly the decoherence rate increases and finally saturates sufficiently close to criticality as the qubit loses sensitivity to dynamics happening at even longer time-scales and length-scales. 
Further, knowing the frequency scale $\omega_0$ (as a function of $d$ and $T$) and the correlation length $\xi$ enables us to predict the scaling of $\langle \phi^2 \rangle$ with distance $d$, time $\tau$, and temperature $T$. 
With this intuition in hand, we next turn to phase transitions in insulating magnets, and derive concrete expressions for $\langle \phi^2 \rangle$ within mean-field theory, as presented in Table~I of the main text.
\newline

\emph{Mean-field theory:} Dynamics at classical phase transitions is typically studied using stochastic dynamical models, which provide a phenomenological description of relaxational dynamics of coarse-grained order parameters \cite{CardyBook,HH69,HHM1,HHM2,HH77}.
This approach is governed by two main principles:  i) at long times, the system relaxes to the appropriate equilibrium Gibbs distribution, and ii) symmetries of the original Hamiltonian are respected in the coarse grained model. This leads to a hydrodynamical theory that includes additional slow ’critical’ modes near the critical point---the correlators of these slow modes are the quantities we are primarily interested in.
Specifying to phase transitions in magnetic insulators, let us consider two different Ising transitions with two different symmetries. 
In both cases, the order parameter is $S_\mathrm{z}$, and on lowering $T$ there is a phase transition from a paramagnet with $\langle S_\mathrm{z} \rangle = 0$ to an Ising ferromagnet with $\langle S_\mathrm{z} \rangle \neq 0$. However, depending on whether the order parameter is microscopically conserved on not (e.g. because there is an additional U$(1)$ symmetry), we have different relaxation dynamics and thus $\langle \phi^2 \rangle$ exhibit distinct behaviors near the critical point. 
Note that, since we consider discrete symmetry breaking, there are no Goldstone modes on the ordered side. As a result, $\langle \phi^2 \rangle$ behaves qualitatively the same when approaching the critical point from either side of the transition.

\emph{Model A:} We first consider the case Ising order parameter is not conserved by the microscopic Hamiltonian (model A in the Halperin-Hohenberg taxonomy \cite{HH77}), an example would be a fully anisotropic exchange Hamiltonian:
\beq H_A = -\sum_{\langle i j \rangle}  J_\mathrm{x} S^\mathrm{x}_i S^\mathrm{x}_j + J_\mathrm{y} S^\mathrm{y}_i S^\mathrm{y}_j + J_\mathrm{z} S^\mathrm{z}_i S^\mathrm{z}_j, \text{ with } J_\mathrm{z} > J_\mathrm{y} > J_\mathrm{x} > 0
\eeq
As alluded to earlier, we write down a stochastic differential equation governing the time-evolution of the coarse grained order parameter $\varphi(\r,t) \sim S^\mathrm{z}_{\r_i}(t)/S$ ($S$ is the magnitude of spin):
\beq
\partial_t \varphi(\r,t) = - \Gamma_0 \frac{\delta F}{\delta \varphi(\r,t)} + \zeta(\r,t), \text{ with } 
F = \int d\r \left[   J \left( \frac{1}{2} (\nabla \varphi)^2  + \frac{r}{2} \varphi^2 + \frac{u}{4} \varphi^4 + ... \right) -  h(\r,t) \varphi(\r,t) \right]
\eeq
where $F$ is an effective Ginzburg-Landau free-energy, $J \approx J_\mathrm{z}$ sets the overall energy scale, $h(\r,t)$ is an external field that couples linearly to the order parameter, and the white noise $\zeta(\r,t)$ (coming from short wavelength, fast fluctuations) satisfies the following condition on the long wavelengths and timescales of our interest:
\beq
\langle \zeta(\r,t) \rangle = 0 \quad \text{ and } \quad \langle \zeta(\r,t) \zeta(\rp,t^\prime) \rangle = 2 T \Gamma_0 \, \delta(\r - \rp) \, \delta(t - t^\prime)
\eeq
Within mean-field theory we set $u =0$, and $r = \xi^{-2} \propto |T - T_\mathrm{c}|$ quantifies the distance from the critical point ($\xi$ being the correlation length that diverges at $T = T_\mathrm{c}$). 
The equation for the $(\q,\omega)$ Fourier mode of $\varphi$ reads:
\beq
-i \omega \varphi(\q,\omega) = -\Gamma_0 J (\q^2 + \xi^{-2}) \varphi(\q,\omega) + \Gamma_0 h(\q,\omega)  + \zeta(\q,\omega), \text{ where } \varphi(\q,\omega) = \int dt \int d\r \, e^{i (\q \cdot \r - \omega t)} \varphi(\r,t)
\eeq
We can now evaluate the dynamic susceptibility $\chi(\q,\omega)$:
\beq
\chi(\q,\omega) = \frac{\partial \langle \varphi(\q,\omega) \rangle}{\partial h(\q,\omega)}\bigg|_{h = 0} = \frac{\Gamma_0}{\Gamma_0 J(\xi^{-2} + \q^2) - i \omega } = \left[\chi^{-1}(\q) - \frac{i\omega}{\Gamma_0}\right]^{-1}, \text{ with } \chi^{-1}(\q) = J(\xi^{-2} + \q^2)
\label{eq:chiModelA}
\eeq
We note that in the static limit $\omega \rightarrow 0$, we recover the static correlator $\chi(\q) \equiv \chi(\q, \omega = 0)$; the uniform static correlator [$\chi_u \equiv \chi(\mathbf{0},0)$] diverges at the critical point, resulting in a gapless mode that enhances the noise. The dynamic structure factor follows from the fluctuation dissipation theorem:
\beq
S(\q,\omega) = \frac{2T}{\omega} \text{Im}[\chi(\q,\omega)] =  \frac{2 T \Gamma_0}{\Gamma_0^2 J^2 (\xi^{-2} + \q^2)^2 + \omega^2}
\label{eq:SModelA}
\eeq
We can in principle find the scaling of $\langle \phi^2 \rangle$ in different regimes by just plugging this into Eq.~(2) in the main text, which we recall here for completeness:
\beq
\langle \phi^2\rangle = \int \frac{d\omega}{2\pi}~W_{\tau}(\omega) \underbrace{\int_0^\infty \frac{dq}{2\pi} ~ W_d(q) S(q, \omega)}_{\mathcal{N}(\omega)}, \text{ where } W_d(q) \propto q^3 e^{-2qd}
\label{eq:phiSqApp}
\eeq
While the general expression for $\langle \phi^2\rangle$ as a function of $d, \tau$ and $T$ can be obtained by numerical integration (analytically, they assume a non-transparent hypergeometric form), here we focus on a few simple asymptotic limits. 
We will find that all these asymptotic scalings can also be obtained by simply substituting the typical frequency scale $\omega_0 = \Gamma_0 J (\xi^{-2} + d^{-2})$, for typical $q \sim d^{-1}$, within the noise spectral density $\N(\omega)$ in our heuristic formula in Eq.~\eqref{eq:AppScR}.

First, consider the limit when $d \gg \xi$, a regime that typically holds away from criticality as the correlation length $\xi$ is small. 
In this case, we can neglect the momentum-dependence of $S(\q,\omega)$ in Eq.~\eqref{eq:SModelA}, as it is well-approximated by its uniform ($\q = 0$) value in the regime where $W_d(q)$ is appreciably large, and exhibits a typical frequency width $\omega_0 \approx \Gamma_0 J \xi^{-2}$. 
Therefore, we can carry out the momentum integral, which yields:
\beq
\langle \phi^2\rangle \approx T \left( \int\frac{ d\omega}{2\pi} \,  \frac{W_\tau(\omega) \Gamma_0}{\Gamma_0^2 J^2 \xi^{-4} + \omega^2} \right) \left( \int \frac{dq}{2\pi} W_d(q) \right) \approx \frac{T \Gamma_0}{d^4}\int\frac{ d\omega}{2\pi} \frac{W_\tau(\omega)}{\Gamma^2 J^2 \xi^{-4} + \omega^2}
\eeq
When performing Ramsey spectroscopy, if the width $\tau^{-1}$ of the frequency filter function is narrow on the scale of $\omega_0$, then we can approximate $W_\tau(\omega) \propto \tau \delta(\omega)$. In contrast, if the width of $W_\tau(\omega)$ is large such that $\tau^{-1} \gg \omega_0$, then we can instead replace $W_\tau(\omega)$ by its $\omega \to 0$ limit of $\propto \tau^2$; here we integrate over the entire noise spectrum and consequently capture the static correlation function:
\beq
\langle \phi^2\rangle \approx \begin{cases} 
\left[ \dfrac{T \tau \xi^2}{J d^4}\right]\dfrac{\xi^2}{\Gamma_0 J}   & \omega_0 \tau \gg 1 \\[15pt]
\left[ \dfrac{T \tau \xi^2}{J d^4}\right] \tau & \omega_0 \tau \ll 1 \end{cases}
\label{eq:phiSqModelASmallxi}
\eeq
Both cases in Eq.~\eqref{eq:phiSqModelASmallxi} are reproduced by our heuristic formula in Eq.~\eqref{eq:AppScR} by substituting $\omega_0 = \Gamma_0 J(d^{-2} + \xi^{-2})$, and taking appropriate limits.

Next, let us consider the critical regime (or small distances) such that $d \ll \xi$. In this case, momentum dependence of $S(\q,\omega)$ needs to be taken into account more carefully. When $\tau$ is large such that $\omega_0 \tau \gg 1$, the dominant contribution from $\N(\omega)$ comes at low frequencies. Defining a dimensionless momentum, $\bar{q} = qd$, we can capture the leading divergence in $\N(\omega)$ as follows:
\beq
\N(\omega) = \frac{2T }{\Gamma_0 d^4 J^2} \int_0^\infty d\bar{q} \, \frac{\bar{q}^3 e^{-2\bar{q}}}{\bar{q}^4 + \left( \omega d^2/\Gamma_0 J \right)^2} \approx  \frac{2T}{\Gamma_0 J^2} \ln \left(\frac{\Gamma_0 J}{\omega d^2} \right) \text{ when } \omega \to 0 
\eeq
Integrating against $W_\tau(\omega)$ now yields an additional logarithmic correction to the heuristic formula in Eq.~\eqref{eq:AppScR}:
\beq
\langle \phi^2\rangle \approx \frac{2T \tau}{\Gamma_0 J^2} \ln \left(\frac{\Gamma_0 J \tau}{ d^2} \right), ~~~ \omega_0 \tau \gg 1 \text{ and } d \ll \xi
\eeq

In the opposite limit $\omega_0 \tau \ll 1$, it is more convenient to switch the order of integrals in Eq.~(\ref{eq:phiSqApp}), and do the frequency integral first after Taylor expanding $W_\tau(\omega)$, whereby we recover the static structure factor at criticality (as expected). 
\beq
\int \frac{d\omega}{2\pi} W_\tau(\omega) S(\q,\omega) \approx \tau^2 \int d\omega \, \frac{\Gamma_0}{\Gamma_0^2 J^2 q^4 + \omega^2} = \frac{2T}{J q^2} = 2 T \chi(\q,0) 
\eeq
The momentum integral can now be straightforwardly evaluated, leading to a result consistent with our heuristic scaling. 
\beq
\langle \phi^2\rangle \approx \frac{T \tau^2}{J d^2}, ~~~ \omega_0 \tau \ll 1 \text{ and } d \ll \xi
\eeq

\emph{Model B:} We next consider the case where the Ising order parameter is conserved by the microscopic Hamiltonian (model B in the Halperin-Hohenberg taxonomy) \cite{CardyBook,HH77}. 
An example is the easy-axis XXZ model on a square lattice, with a Hamiltonian $H_B$ that conserves total $S^\mathrm{z}$:
\beq H_B = -\sum_{\langle i j \rangle} J_\parallel S^\mathrm{z}_i S^\mathrm{z}_j + J_\perp ( S^\mathrm{x}_i S^\mathrm{x}_j +  S^\mathrm{y}_i S^\mathrm{y}_j), \text{ with } J_\parallel > J_\perp > 0
\eeq

Like model A, we can write down an effective Landau-Ginzburg free-energy functional $F$ for $\varphi(\r,t) \sim S^\mathrm{z}_{\r_i}(t)/S$, which takes an identical form.  
Accordingly, the dynamical evolution equation for the Fourier mode $(\q,\omega)$ of the order parameter $\varphi$ is given by:
\beq
-i \omega \varphi(\q,\omega) = -J \Gamma(\q) (\q^2 + \xi^{-2}) \varphi(\q,\omega) + \Gamma(\q) h(\q,\omega)  + \zeta(\q,\omega),
\eeq
However, since $\varphi(\q = 0, t) = \int d\r \, \varphi(\r,t)$ is conserved, this implies that $\Gamma(\q = 0) = \Gamma_0 = 0$, in sharp contrast to model A which has $\Gamma_0 \neq 0$. 
Assuming that the interactions are short-ranged, we expect $\Gamma(\q)$ to be analytic in $\q$ and therefore can write it as $\Gamma(\q) = \sigma_s \q^2$ to lowest order in momentum \cite{CardyBook,HH77}. 
Therefore, we should replace $\Gamma_0$ by $\Gamma(\q) = \sigma_s \q^2$ in the expression for $\chi(\q,\omega)$ in Eq.~\eqref{eq:chiModelA}, leading to:
\begin{align}
\chi(\q,\omega) = \left[-\frac{i \omega}{\sigma_s \q^2} + J(\q^2 + \xi^{-2})\right]^{-1} =& \frac{\chi_u D_s \q^2}{- i \omega + D_s \q^2 + J \sigma_s \q^4}, \label{eq:chiforDiffusion} \\
\text{ where } \chi_u = \lim_{\q \to 0} \lim_{\omega \to 0} \chi(\q,\omega) = \frac{\xi^{2}}{J}, &\quad D_s \equiv \frac{\sigma_s}{\chi_u} = \frac{\sigma_s}{J\xi^2}
\end{align}
Thus, if we neglect the higher order momentum correction in the denominator in Eq.~\eqref{eq:chiforDiffusion}, we recover the diffusive form of the correlator with diffusion constant $D_s = \sigma_s \chi_u^{-1}$ (Einstein relation). 
Further, $D_s$ goes to zero at the critical point, indicative of the critical slowing down, and the higher order terms of $\Gamma(\bm{q})$ become important. 
The dynamic structure factor therefore can be written as:
\beq
S(\q,\omega) = \frac{2T}{\omega} \text{Im}[\chi(\q,\omega)] = 
\frac{2 T \sigma_s \q^2}{(\sigma_s \q^2)^2(\xi^{-2} + \q^2)^2 + \omega^2} 
\label{eq:SModelB}
\eeq
Once again, the general $\langle \phi^2 \rangle$ may be found by numerical integration, and we focus on the asymptotic limits which are analytically amenable. When $d \gg \xi$, we can approximate the structure factor as:
\beq
S(\q,\omega) \approx \frac{2 T \chi_u D_s \q^2}{\omega^2 + (D_s \q^2)^2}
\eeq
The typical frequency scale in this regime is given by $\omega_0 = D_s d^{-2}$ (using typical $q \sim d^{-1}$).
In the large time limit, i.e, $\omega_0 \tau \gg 1$, the qubit is again sensitive to $S(\q,\omega = 0) = \frac{2 T \chi_u}{D_s \q^2}$. 
In contrast, in the small time limit, i,e, $\omega_0 \tau \ll 1$, the qubit can sense the entire spectral weight of the diffusive mode as the filter function width ($\tau^{-1}$) is much larger than the typical spread $\omega_0$ of the structure factor $S(q = d^{-1},\omega)$.
The integral over all frequencies corresponds to measuring the static structure factor $S(\q) = \int \frac{d\omega}{2\pi} S(\q,\omega)$. These considerations lead to the following expression for $\langle \phi^2 \rangle$ in the diffusive regime. 
\beq
\langle \phi^2\rangle \approx \begin{cases} 
\left[\dfrac{T \tau \xi^2}{d^2}\right] \dfrac{1}{D_s} &\omega_0 \tau \gg 1 \\[15pt] 
\left[\dfrac{T \tau \xi^2}{d^2}\right] \dfrac{\tau}{d^2} & \omega_0 \tau \ll 1 \end{cases}
\eeq
We note that while the small time limit is identical to model A, the large time limit is different as we are sensitive to the dynamics (set by $\omega_0$) which differ significantly between the two models. 
Further, both limits are in accordance with our heuristic formula in Eq.~\eqref{eq:AppScR}.

Close to criticality, the diffusion coefficient vanishes, $D_s \to 0$, and the behavior of spin-correlations is no longer diffusive.
Rather, from Eq.~\eqref{eq:SModelB}, the typical frequency scale is now given by $\omega_0 = \sigma_s d^{-4}$ (for typical $q \sim d^{-1}$). 
Once again in the large $\tau$ limit, i.e, $\omega_0 \tau \gg 1$, the dominant contribution from $\N(\omega)$ comes at low frequencies. Defining $\bar{q} = qd$, we can capture the leading divergence in $\N(\omega)$ as follows:
\beq
\N(\omega) = \frac{2T }{\sigma_s d^6} \int_0^\infty d\bar{q} \, \frac{\bar{q}^5 e^{-2\bar{q}}}{\bar{q}^8 + \left( \omega d^4/\sigma_s \right)^2} \approx  \frac{T}{\sqrt{\sigma_s \omega}} \text{ when } \omega \to 0 
\eeq
This $1/\sqrt{\omega}$ divergence in the noise spectral density at low-frequency (which corresponds to a $1/\sqrt{t}$ decay of the correlations at large time) leads to (with $\bar{\omega} = \omega \tau$):
\beq
\langle \phi^2\rangle \approx \int d\omega \frac{4 \sin^2(\omega \tau/2)}{\omega^2} \frac{T}{\sqrt{\sigma_s \omega}} = \frac{T}{\sqrt{\sigma_s}}\tau^{3/2} \left( \int d\bar{\omega} \frac{4 \sin^2(\bar{\omega}/2)}{\bar{\omega}^{5/2}} \right) \approx  \frac{T}{\sqrt{\sigma_s}}\tau^{3/2}, ~~~ \omega_0 \tau \gg 1 \text{ and } d \ll \xi
\eeq

Note that, once gain, in the short time limit is identical to model A as we are insensitive to dynamics. 
By reversing the order of frequency and momentum integration (as we did for model A), we find that:
\beq
\langle \phi^2\rangle \approx \frac{T \tau^2}{d^2}, ~~~ \omega_0 \tau \ll 1 \text{ and } d \ll \xi
\eeq
This concludes our discussion of Ramsey sequence, and the above results are summarized in the first two rows of Table~I of the main text. 

\emph{General scaling functions:} The most general scaling for $\langle \phi^2 \rangle$, beyond mean-field approximation, can be written down by analyzing the scaling behavior of the susceptibility in the vicinity of $T_\mathrm{c}$ in terms of a scaling function $\chi_{sc}$:
\beq
\chi(\q,\omega) = q^{-2 + \eta} \, \chi_{sc}\left( \omega q^{-z}, q \xi \right)
\eeq
Leveraging the fluctuation dissipation theorem, we can write down a scaling function $S_{sc}(\q,\omega)$ for the dynamic structure factor:
\begin{align}
S(\q,\omega) &= \frac{2T}{\omega} \text{Im}[\chi(\q,\omega)] \\
&=  \frac{2T}{\omega}  q^{-2 + \eta} \text{Im}[\chi_{sc}(\omega q^{-z}, q \xi)] \\
&=T q^{-2 + \eta - z} \tilde{S}_{sc}(\omega q^{-z}, q \xi) \\
&= T q^{-2 + \eta - z} S_{sc}(\omega \xi^{z}, q \xi) 
\end{align}
where we have re-defined the scaling form $\text{Im}[\chi_{sc}(\omega q^{-z}, q \xi)] = \omega q^{-z} \tilde{S}_{sc}(\omega q^{-z}, q \xi)$ in the penultimate step, and used $\omega q^{-z} = (\omega \xi^z)\times(q\xi)^{-z}$ in the last step to re-write the scaling function using a convenient set of variables.  Plugging this into Eq.~(\ref{eq:phiSqApp}) and non-dimensionalizing the frequency and momentum integral in terms of $\bar{\omega} = \omega \tau$ and $\bar{q} = qd$ respectively, leads to:
\beq
\langle \phi^2 \rangle = \frac{T \tau}{d^{2 + \eta - z}} \int \frac{d\bar{\omega}}{2\pi} \frac{4 \sin^2(\bar{\omega})}{\bar{\omega}^2} \int \frac{d\bar{q}}{2\pi}\, \bar{q}^3 e^{-2 \bar{q}} S_{sc} (\bar{\omega}  \xi^{z}/\tau, \bar{q} \xi/d) = \frac{T \tau}{d^{2 + \eta - z}} F\left( \frac{\tau}{\xi^z}, \frac{d}{\xi} \right)
\eeq
which is exactly Eq.~(4) in the main text. 
\newline

\begin{figure}
    \centering
    \includegraphics[width = \textwidth]{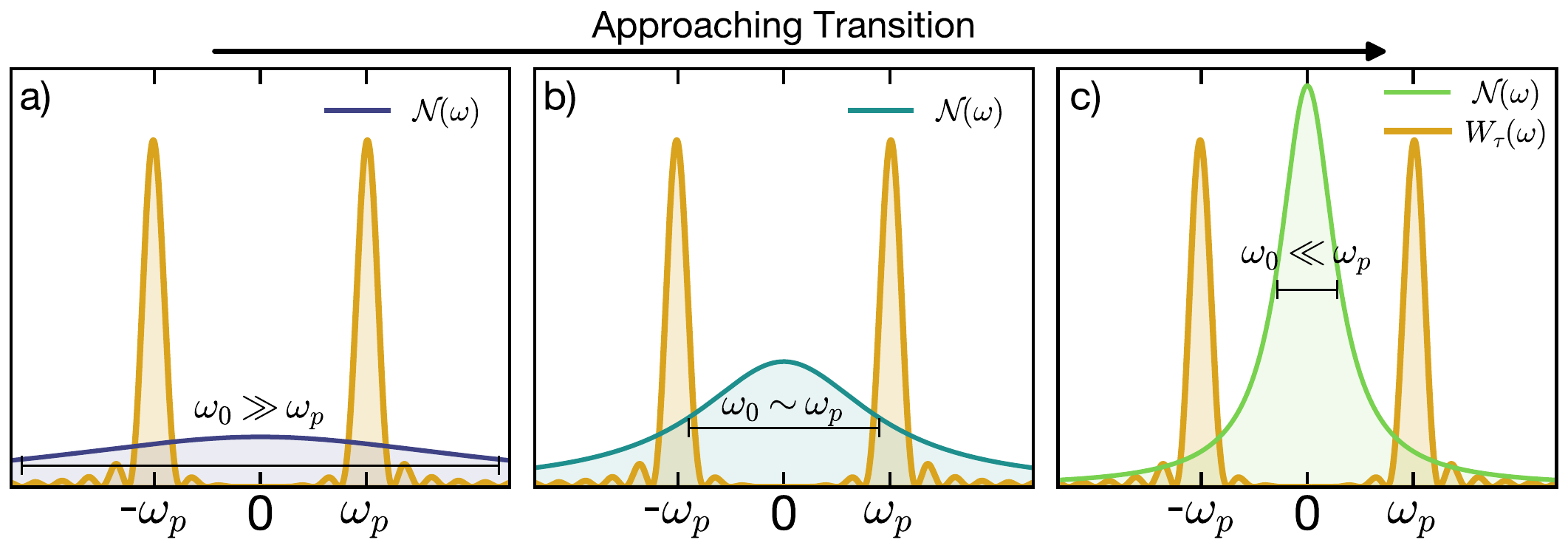}
    \caption{Schematic depiction of the noise spectral density $\N(\omega)$ and the filter function $W_{\tau}(\omega)$ as we approach the critical point, indicating that the maximum overlap happens close to, but not exactly at the critical point (when $d$ is large).
    }
    \label{fig:NcrossoverEcho}
\end{figure}

\emph{CPMG:} 
In this section, we discuss the signatures of criticality that can be obtained when using more intricate frequency filter function $W_\tau(\omega)$ such as the CPMG pulse sequences (see Fig.~\ref{fig:CPMG} for a schematic).
While the exact analytic expressions for the filter function can be found in Eq.~\eqref{eq:FreqFilter}, we will find it more convenient to focus on the limit of a large number of pulses, derived in Eq.~\eqref{eq:largeNFreqFilter}.
In this limit, the frequency filter function takes the form of a sum of Dirac-delta functions, with the most prominent peak at $\omega_p = N \pi/\tau$ (see Fig.~\ref{fig:NcrossoverEcho}).
Therefore, the characteristic frequency of measurement changes from $\omega = 0$ for Ramsey to $\omega \approx \omega_p$ for CPMG.
The elimination of the dc component makes the noise profile sensitive to dynamics of sources even in the short measurement-time limit (typical noise frequency $\omega_0 \ll \omega_p$), while leaving the opposite limit of large measurement time essentially unchanged.
Using that the noise spectral density takes a Lorentzian form [i.e., $\N(\omega) \propto \frac{\omega_0}{\omega_0^2 + \omega^2}$ close to the critical point as discussed the previous section], we can derive an explicit analytic form for $\langle \phi^2 \rangle$ for CPMG sequences:
\begin{align}
\langle \phi^2 \rangle &\propto  \tau \int \frac{d\omega}{2\pi} \left[ \sum_{n=0}^\infty \frac{\delta\left[\omega - (2n+1)\omega_p \right]}{(2n+1)^2} \;  \right] \mathcal{N}(\omega) = \frac{\tau}{2\pi} \sum_{n=0}^{\infty} \frac{\mathcal{N}\left((2n+1)\omega_p\right) }{(2n+1)^2} \; \\
&= \frac{\tau}{2\pi} \sum_{n=0}^{\infty} \frac{1}{(2n+1)^2} \; \left[ \frac{\omega_0}{\omega_0^2 + (2n+1)^2 \omega_p^2} \right] = \frac{\pi \tau}{16 \omega_0} \left[ 1 - \frac{\tanh(\pi \omega_0/2\omega_p)}{\pi \omega_0/2\omega_p} \right]
\label{eq:PhiSqCPMGLargeN}
\end{align}

Combining asymptotic limits of Eq.~\eqref{eq:PhiSqCPMGLargeN} with the intuition that the distance scaling of the noise is not affected by the choice of frequency filter function, we can write down the following asymptotic expressions for the noise:
\beq
\langle \phi^2 \rangle \approx \begin{cases} \dfrac{\tau}{\omega_0 d^2} \times \rm{min}\left(1, \dfrac{\xi^2}{d^2} \right) &\omega_0 \gg \omega_p  \\[15pt]
\dfrac{\tau}{\omega_0 d^2} \left( \dfrac{\pi\omega_0}{2\omega_p}\right)^2\times \rm{min}\left(1, \dfrac{\xi^2}{d^2} \right) &\omega_0 \ll \omega_p
\end{cases}
\label{eq:PhiSqCPMGlimits}
\eeq
Away from criticality, the noise spectrum $\mathcal{N}(\omega)$ is quite flat at low-frequencies ($\omega_0 \gg \omega_p$). 
Therefore, the qubit probe is essentially sensitive to $\mathcal{N}(\omega \approx 0)$ (see Fig.~\ref{fig:NcrossoverEcho}a). 
Thus, in this regime CPMG provides the same information as Ramsey spectroscopy. 
As we approach the critical point, $\omega_0$ decreases due to critical slowing down, and consequently $\langle \phi^2 \rangle$ increases.
This increase continues until $\omega_0 \sim \omega_p$, where $\langle \phi^2 \rangle$ is maximized (see Fig. \ref{fig:NcrossoverEcho}b).
Upon reducing $\omega_0$ further, $\mathcal{N}(\omega)$ becomes sharply peaked and the overlap of $\mathcal{N}(\omega)$ with the CPMG filter function $W_\tau(\omega)$ decreases (see Fig.~\ref{fig:NcrossoverEcho}c), resulting in a corresponding decrease of $\langle \phi^2 \rangle$.
However, given that typical energy-scales in electronic systems are much larger than $\omega_p$ ($\sim$ mK), tuning a system to the regime where $\omega_0 \ll \omega_p$ might be out of reach for current experimental setups.
Therefore, even for CPMG pulse sequences, we generically expect to observe an increase in $\langle \phi^2 \rangle$ as we tune towards the critical point.

Next, we turn to explicit expressions for $\langle \phi^2 \rangle$ in asymptotic regimes for thermal transitions.
To start, we focus on the limit of $\omega_0 \ll \omega_p$.
In this regime, $\langle \phi^2 \rangle$ is sensitive to the tail of the noise spectral density $\mathcal{N}(\omega)$, which contains information about critical dynamics (see Fig.~\ref{fig:NcrossoverEcho}c).  
Therefore, we can use CPMG spectroscopy in this regime to distinguish between models in different dynamic universality classes. 
Taking up our previous examples, we find that for model A (Ising model with no conservation law):
\beq
\langle \phi^2 \rangle \approx \frac{T \tau \Gamma_0}{\omega_p^2 d^4}
\eeq
For model B (Ising transition with a conserved order parameter), the dependence of $\langle \phi^2 \rangle$ on qubit-sample distance $d$ is distinct:
\beq
\langle \phi^2 \rangle \approx \frac{T \tau \sigma_s}{\omega_p^2 d^2}
\eeq
As alluded before, in the opposite limit ($\omega_0 \gg \omega_p$), the scaling of $\langle \phi^2 \rangle$ for CPMG pulse sequences is identical to Ramsey spectroscopy.

\section{Noise at quantum phase transitions}
In this section, we expand on the discussion of decoherence signatures of quantum phase transitions, and provide derivations of the scaling of $\langle \phi^2 \rangle$ for some representative examples. As discussed in the main text, dynamics at quantum phase transitions can be studied using deterministic evolution by the quantum Hamiltonian itself, without resorting to a stochastic description. Specifying to insulators with spin-degrees of freedom, we derive $\langle \phi^2 \rangle$ for the Ramsey sequence using the low-energy dynamics in the vicinity of a quantum critical point for two illustrative transitions (both in two spatial dimensions): (i) N\'{e}el to quantum paramagnet (valence bond solid) transition driven by geometric frustration, and (ii) Field-driven ferromagnet to paramagnet transition in a transverse field Ising model (TFIM). 
While the former was discussed in the main text, the latter is important as an example of a transition without any conservation law and therefore serves as an analogue to the anisotropic Ising transition (model A) discussed in the context of classical/thermal phase transitions. 
\newline 

\emph{N\'{e}el-paramagnet transition}: We consider a Heisenberg Hamiltonian on a square lattice: 
\beq 
H =  \sum_{i \neq j} J_{ij} \S_i \cdot \S_{j}, \text{ with short-ranged } J_{ij} > 0.
\eeq 
In presence of only nearest neighbor exchange, the ground state of $H$ is well-known to be a N\'{e}el antiferromagnet. 
As we tune the couplings $J_{ij}$ to increase frustration, quantum fluctuations drive a transition to paramagnetic phase. The low-energy properties of this transition is captured by the O(3) non-linear sigma model, and has been discussed extensively in the literature \cite{CHN1,CHN2,THC89,CSY94,SY92,ssbook}; here we follow Ref.~\onlinecite{ssbook}. 
The O(3) non-linear sigma model for the order parameter field $\n(\r_i) \sim (-1)^{i_{x} + i_{y}} \S_{\r_i}/S$ is described by the action:
\beq
S = \int dt \int d\r \, \mathcal{L}(\r,t), ~~ \mathcal{L} =  (\partial_t \n)^2 - c^2 (\nabla \n)^2 \text{ with } \n^2 = 1
\eeq
The length constraint introduces non-linearity in the otherwise linear action. In order to formulate the quantum Hamiltonian for the effective theory, it is helpful to introduce the conjugate angular momentum $\L$ which generates rotations of $\n$ i.e, $[L_\alpha, n_\beta] = i \varepsilon_{\alpha \beta \gamma} n_\gamma$; the fluctuations of $\L$ can be interpreted as the `ferromagnetic' fluctuations. A lattice-regularized Hamiltonian is therefore given by:
\beq
H = \frac{J \lambda}{2} \sum_i \L_i^2 - J \sum_{\langle ij \rangle} \n_i \cdot \n_j - \H \cdot \sum_{i} \L_i
\eeq
where we have introduced an external field $\H$ which couples to the net conserved angular momentum $\L_{\rm tot} \equiv \sum_i \L_i$. 
For $\H = 0$, this Hamiltonian has an ordered antiferromagnetic phase for small $\lambda$ with $\langle \n \rangle \neq 0$, and a quantum-disordered phase at large $\lambda$ with $\langle \n \rangle = 0$.

Since the typical sample-probe distance $d \gg a$ (lattice spacing), it follows that qubit probe is not sensitive to short-wavelength spin-fluctuations at the scale of the lattice spacing (the momentum filter function $W_d(q)$ is exponentially small for $q \sim a^{-1}$). 
Therefore, the qubit probe cannot directly access the order parameter fluctuations $\langle n_\mu(\r,t) n_\nu(0,0)\rangle$. 
Rather, we detect the slowly varying correlations of the uniform spin-density, which corresponds to $C_{\mu \nu}(\r, t) \equiv \langle L_\mu(\r, t) L_\nu(0,0) \rangle$.
It is more convenient to study the dynamic susceptibility $\chi_{\mu \nu}(\q, \omega)$, defined as follows \cite{ssbook}):
\beq
\chi_{\mu \nu}(\q, \omega) &\equiv& i\int_0^\infty dt \int d\r \, C_{\mu \nu}(\r, t) e^{i (\omega t - \q \cdot \r)}
\eeq
At small $\q$ and $\omega$, at any finite temperature there is unbroken spin-rotation invariance (no order). 
Since $\L(\r,t)$ is a conserved density, at long times the susceptibility takes a diffusive form:
\beq
\chi_{\mu \nu}(\q, \omega) = \delta_{\mu \nu}\chi(\q, \omega), \text{ with } \chi(\q, \omega) =  \frac{\chi_{u} D_s \q^2}{- i \omega + D_s \q^2}
\eeq
Of course, this hydrodynamic form requires that the typical $\omega$ satisfies $\omega t_\mathrm{c} \ll 1$, where $t_\mathrm{c}$ is the microscopic collision time: this approximation is well-justified because the typical time-scale $\tau$ in the frequency filter function $W_\tau(\omega)$ is much longer than $t_\mathrm{c}$. 
We use the fluctuation-dissipation theorem to extract the dynamic spin structure factor from the susceptibility \cite{ssbook,coleman_2015}.
This leads to the following form for $\langle \phi^2 \rangle$ in Ramsey: 
\beq
\langle \phi^2 \rangle \approx \begin{cases} \left[\dfrac{T\tau\chi_u}{d^2}\right] \dfrac{\tau}{d^2} & \omega_0 \tau \ll 1 \\[12pt]
\left[\dfrac{T\tau\chi_u}{d^2}\right] \dfrac{1}{D_s} &  \omega_0 \tau \gg 1 \end{cases}
\label{eq:T2sO3}
\eeq
where $\omega_0 \sim D_s/d^2$ is assumed to satisfy $\omega_0 t_\mathrm{c} \ll 1$. 
The spin-diffusion coefficient $D_s$ and the uniform static susceptibility $\chi_{u}$ can both be written down in terms of scaling functions of $\delta/T$, where $\delta$ refers to an intrinsic energy scale in the system, which is set by stiffness $\rho_s$ for $\lambda < \lambda_\mathrm{c}$ (ordered side) and single particle gap $\Delta$ for $\lambda > \lambda_\mathrm{c}$ (paramagnetic side). The behavior of these scaling functions therefore dictate $\langle \phi^2 \rangle$, which are strongly constrained by symmetry considerations; we turn to these next.

First, we consider the uniform static susceptibility $\chi_u$, which can be written as the second derivative of the free energy density $f[\H] = - (T/V) \ln[Z[\H]]$ with respect to the external field, i.e,
\beq
f[\H] = f[\H = 0] - \frac{1}{2} \chi_{u,\alpha \beta}  H_\alpha  H_\beta + ...
\eeq
Since the free energy density has dimensions $d + z$ in $d$-dimensional space, and $\H$ causes $\n$ to Larmor precess ($\partial_\tau \n \rightarrow \partial_\tau \n - i \H \times \n$) and therefore has fixed dimension $-z$, so we have $[\chi_u] = d - z$. Therefore, the scaling form for $\chi_u$ is (we now restrict to $d = 2$):
\beq
\chi_u = \frac{T}{c^2} \Phi_{u,\pm} \left( \frac{\delta}{T} \right)
\eeq
where we have used $[T] = z$, $[c] = z - 1$, $\Phi_{u,\pm}$ are dimensionless scaling functions, and $\delta = \rho_s$ for $\lambda < \lambda_\mathrm{c}$ (ordered side) and $\delta = \Delta$ for $\lambda > \lambda_\mathrm{c}$ (paramagnetic side).
Similarly, because diffusion involves transport of a conserved density, the diffusion constant $D_s$ does not pick up an anomalous dimension, i.e, $[D_s] = z - 2$. 
Therefore, as a function of temperature $T$ and energy-scale $\delta$, $D_s$ has a scaling form given by:
\beq
D_s = \left( \frac{c^2}{T} \right) \, \Phi_{D_s,\pm} \left( \frac{\delta}{T} \right)
\eeq
Combining this information, we have the following scaling form for the noise (setting $z = 1$, the case for general $z$ is quoted in Table~I of the main text):
\beq
\langle \phi^2 \rangle \approx \begin{cases} \left[ \dfrac{T^2 \tau}{c^2 d^2} \Phi_{u,\pm}  \left( \dfrac{\delta}{T} \right) \right]   \dfrac{\tau}{d^2} & \omega_0 \tau \ll 1 \\[15pt]
\left[ \dfrac{T^2 \tau}{c^2 d^2} \Phi_{u,\pm}  \left( \dfrac{\delta}{T} \right) \right]  \dfrac{T}{c^2 \, \Phi_{D_s,\pm}  \left( \delta/T \right) } & \omega_0 \tau \gg 1
\end{cases}
\eeq

Now let us discuss the different universal regimes where the functions $\Phi_{u}$ and $\Phi_{D_s}$ are known functions of the ratio $\delta/T$. 
First, let us consider the antiferromagnetic regime $\lambda < \lambda_\mathrm{c}$ and low T, i.e, $T \ll \delta = \rho_s$. 
In this regime, the correlation length $\xi \sim \frac{c}{T} e^{ 2 \pi \rho_s/T } $ is exponentially large at low T \cite{CHN1,CHN2,THC89}. 
This implies that up to momenta of order $q \approx \xi^{-1}$, bosonic spin-wave excitations have very high occupancy:
\beq
n_B(q \approx \xi^{-1}) = (e^{\beta c \xi^{-1}} - 1)^{-1} \approx \xi/(\beta c) = e^{2 \pi \rho_s/T} \gg 1
\eeq
Therefore, in this regime we can describe the system using quasi-classical thermal spin waves. If $q \xi \gtrsim 1$, the spin-waves are slowly damped, and therefore the behavior is analogous to a classical Heisenberg antiferromagnet with renormalized parameters. Therefore, for $d \lesssim \xi$ the system appears ordered and the spin-spin correlations are governed by weakly damped spin-waves. 
In contrast, if $d \gtrsim \xi$, i.e, we are probing the lowest momentum scales ($q \xi \lesssim 1$), then the spin-waves get heavily damped.
Therefore, for a large qubit-probe distance $d$, we instead measure diffusive behavior in the correlation function, as discussed in Eq.~(\ref{eq:T2sO3}). 
The values of $\chi_u$ and $D_s$ are given by (for $T \ll \rho_s$) \cite{ssbook}:
\beq
\chi_u = \frac{T}{c^2} \left[ \frac{2\rho_s}{3T} + \frac{1}{3\pi} + O
\left( \frac{T}{\rho_s} \right) \right], \text{ and } D_s \approx \frac{c^2}{T} \left( \sqrt{\frac{T}{\rho_s}}  e^{2 \pi \rho_s/T} \right)
\label{eq:chiDslowTOrdered}
\eeq
Since the diffusion constant diverges as $T \to 0$ and the uniform susceptibility $\chi_u$ remains finite, $\langle \phi^2 \rangle$ goes to zero exponentially in $\rho_s/T = \delta/T$ in the ordered phase. 
This can be explicitly seen by plugging these expressions for $\chi_u$ and $D_s$ into Eq.~\eqref{eq:T2sO3}:
\beq
\langle \phi^2 \rangle \approx \frac{\tau T^3}{d^2 c^4} \left(\frac{\rho_s}{T} \right)^{3/2} e^{-2\pi \rho_s/T}
\eeq
which is presented in the last row of Table~I in the main text.
We emphasize that observing diffusive behavior of spin-correlations requires that the sample-probe distance $d$ to be much larger than the correlation length $\xi$, which may not hold for very low temperatures.  

Next, we discuss the quantum critical region where temperature is much larger than the stiffness or gap scale ($T \gg \delta$). 
This is the so-called incoherent regime, where the only relevant energy scale is $T$. In this case, both the susceptibility and diffusion constant are set purely by the temperature \cite{ssbook}:
\beq
\chi_u = \frac{\sqrt{5}}{\pi} \ln\left( \frac{\sqrt{5}+1}{2} \right) \frac{T}{c^2}, \text{ and } D_s \approx 0.3/\chi_u \sim \frac{c^2}{T}
\eeq
Note that as $T$ approaches $\rho_s$ in Eq.~(\ref{eq:chiDslowTOrdered}), there is a crossover from the low-T large stiffness regime to the critical regime in both $\chi_u$ and $D_s$. 
In this regime, using Eq.~\eqref{eq:T2sO3}, we find the following expression for scaling of the noise in the long time ($\omega_0 \tau \gg 1)$ limit:
\beq
\langle \phi^2 \rangle \approx \frac{\tau T^3}{d^2 c^4}
\eeq
which is quoted in Table~I of the main text. 
In the short time limit ($\omega_0 \tau \ll 1$), we note using Eq.~\eqref{eq:T2sO3} 
the noise will scale as  $T^2$ as it is insensitive to the diffusive dynamics. 
Importantly, these temperature scaling behaviors will hold right down to $T = 0$ if we remain exactly at the critical point $\lambda = \lambda_\mathrm{c}$.  

Finally, we discuss the behavior of $\chi_u$ and $D_s$ in the low T paramagnetic regime ($\lambda > \lambda_\mathrm{c}$). 
The ground state is the product state $\prod_{i}\ket{\ell = 0,m_\ell =0}_i$ with zero net angular momentum, and the excitations at low temperature are a dilute gas of gapped bosonic `spin-flips' ($\ell = 1, m_\ell= 0,\pm 1$). 
Because spin-flips have a gap $\Delta$, the uniform susceptibility $\chi_u$ is exponentially suppressed at low $T$. 
However, the spin diffusion constant $D_s$ is still large. 
This can be understood using the Einstein relation $D_s = \sigma_s \chi_u^{-1}$. 
The density of thermally excited quasiparticles are exponentially small, but their scattering time is also exponentially large, resulting in a finite spin-conductivity $\sigma_s$ at low T (with at most a power law divergence in $T$). 
Further, $\chi_u^{-1}$ diverges at low $T$ since the uniform susceptibility $\chi_u$ goes to zero at low temperatures.
Therefore, we expect $D_s = \sigma_s \chi_u^{-1}$ to also diverge at low T, and thus the noise in this regime is uniformly exponentially suppressed. 
For completeness, we provide the expressions for $D_s$ and $\chi_u$ in this regime \cite{ssbook}:
\beq
\chi_u = \frac{\Delta}{\pi c^2} e^{- \Delta/T}, \text{ and } D_s \sim \frac{\ln(\Delta/T)^2}{\chi_u} = \frac{\pi c^2 \ln(\Delta/T)^2 e^{\Delta/T}}{\Delta}
\eeq 
Therefore, we see that in the long-time limit, $\langle \phi^2 \rangle \sim \chi_u/D_s$ goes to zero as $e^{- 2 \Delta/T}$ in the quantum paramagnet at low temperatures as a clear consequence of the spectral gap, as quoted in Table~I in the main text.  
\newline

\emph{Field-driven ferromagnet-paramagnet transition:}
Next, we provide details for the scaling of $\langle \phi^2 \rangle$ for the transverse field Ising model (TFIM) on a square lattice. This differs in two important aspects from the Heisenberg model discussed earlier: (i) The broken symmetry is discrete and there is a spectral gap (above the ground state) on both sides of the critical point, and (ii) The qubit is directly sensitive to the order parameter $\varphi \sim S^\mathrm{z}$, which is not conserved and therefore displays anomalous scaling. This leads to important differences in the scaling of $\langle \phi^2 \rangle$, as we show below.

We start by noting the Hamiltonian for the TFIM, with a dimensionless parameter $\lambda$ controlling the strength of the applied transverse field:
\beq
H = J \sum_{\langle ij \rangle} S^\mathrm{z}_i S^\mathrm{z}_j + J \lambda \sum_{i} S_x
\eeq
While the TFIM is exactly solvable by fermionization in one spatial dimension, and the correlation functions and crossover behavior can be exactly computed \cite{ssbook}; on a square lattice this is no longer the case. However, we can still appeal to the low-energy field theory to extract detailed information about dynamics near the phase transition for the two-dimensional case.
\beq
S = \int dt \int d\r \, \mathcal{L}(\r,t), \; \text{ with } \mathcal{L} = (\partial_t \varphi)^2 -  c^2 (\nabla \varphi)^2 -[ r \varphi^2 + U \varphi^4]
\eeq
Interactions renormalize the value of $r$ to $\lambda - \lambda_\mathrm{c}$, which vanishes at the critical point $\lambda = \lambda_\mathrm{c}$. 
As alluded to earlier, the symmetry broken ($\varphi \rightarrow - \varphi$) is discrete, which means that there exists a gap on both sides of the critical point. 
Therefore, the noise at low T will be exponentially suppressed on both sides of the critical point, but will be large in the quantum critical region. 
Generally, the spectral gap $\Delta$ will scale away from the critical point as:
\beq
\Delta_\pm \propto |\lambda - \lambda_\mathrm{c}|^{\nu z}
\eeq
where we used $\Delta_\pm$ to indicate that the gaps on the two sides of the transition are different.
Secondly, there is no conservation law in the Ising Hamiltonian, which means that spin-correlation functions no longer take a diffusive form, and $\varphi$ can pick up an anomalous dimension $\eta$. 
The probe measures spin correlations and is therefore sensitive to this anomalous dimension. Knowing that $\chi(q,\omega \to 0) \to q^{-2 + \eta}$, we can write down a scaling form for the susceptibility using the dynamical critical exponent $z$, in terms of dimensionless variables $cq/T^{1/z}, \omega/T$ and $\Delta_\pm/T$:
\beq
\chi(q,\omega) = \frac{1}{T^{(2-\eta)/z}} \Phi_{\pm}\left( \frac{cq}{T^{1/z}},\frac{\omega}{T},\frac{\Delta_{\pm}}{T} \right)
\eeq
Plugging this into Eq.~(\ref{eq:phiSqApp}), just as we did for the classical Ising model, leads to the following scaling form for the $\langle \phi^2 \rangle$, in terms of universal scaling functions $\Psi_{\pm}$ (or $\tilde{\Psi}_\pm$):
\beq
\langle \phi^2 \rangle = \frac{\tau}{T^{(2-\eta)/z}d^4} \Psi_{\pm}\left( T \tau, \frac{T^{1/z}d}{c}, \frac{\Delta_{\pm}}{T} \right) = T^{(2 + \eta - z)/z} \tilde{\Psi}_{\pm}\left(T \tau, \frac{T^{1/z}d}{c}, \frac{\Delta_{\pm}}{T} \right)
\label{eq:IsingScaling2pt}
\eeq
For the Ising transition in two spatial dimensions, analytic scaling functions at finite temperature are not available in the literature to the best of our knowledge. 
However, we can appeal to the spectral gap on both sides of the transition to argue that $\langle \phi^2 \rangle$ is exponentially suppressed in $\Delta_\pm/T$ in the low temperature regime. 
In contrast, in the quantum critical regime where temperature is only relevant energy scale, we can use the following phenomenological form:
\beq
\chi(\q, \omega) = \frac{\chi(0,0)}{1 - i \omega/\Gamma  + q^2 \xi^2}, \text{ where } \chi(0,0) \sim T^{(-2 + \eta)/z}, ~~ \Gamma \sim T \text{ and } \xi \sim \frac{c}{T^{1/z}}
\label{eq:QCchiTFIM}
\eeq
This is be motivated by noting that at $\lambda = \lambda_\mathrm{c}$, the static uniform susceptibility $\lim_{q \to 0}\chi(q,0)$ only diverges at $ T = 0$, but remains finite at non-zero T. Since $\chi(q,0) \sim q^{-2 + \eta}$ at $T = 0$ and must obey a scaling relation of the form $\chi(q,0) = q^{-2 + \eta} \chi_{sc}(q T^{1/z})$ at finite temperature, this implies that we can write $\chi(0,0) \sim T^{(-\eta + 2)/z}$. Further, noting that we expect a finite relaxation rate $\Gamma$ for the order-parameter mode $\varphi(q,t)$ towards equilibrium even at $q = 0$ (as the Ising order parameter is not-conserved), and that finite $q$ corrections are expected to be analytic at non-zero $T$ when $\lambda \approx \lambda_\mathrm{c}$; we postulate the form presented in Eq.~(\ref{eq:QCchiTFIM}). 
Lastly, since the only energy-scale in the quantum critical regime is $T$, we must have $\Gamma \sim T$, and $\xi \sim c/T^{1/z}$ (which corresponds to a smaller correlation length at larger $T$ due to thermal fluctuations). 
A similar form is found to be an excellent approximation to the exact solution for low-frequency dynamics of the one-dimensional TFIM \cite{ssbook,SS96}; our arguments show that this should be true for the TFIM in two spatial dimensions as well. 
Plugging the phenomenological functional form of the susceptibility in Eq.~\eqref{eq:QCchiTFIM} into Eq.~(\ref{eq:phiSqApp}), we finally obtain the scaling of $\langle \phi^2 \rangle$ in the experimentally relevant limit of low temperatures:
\beq
\langle \phi^2 \rangle &\approx& \frac{2 T \chi(0,0)}{\Gamma} \int \frac{d\omega}{2\pi} W_\tau(\omega) \int dq \, \frac{q^3 e^{-2qd}}{(1+ q^2 \xi^2)^2 + (\omega/\Gamma)^2}  \approx  \frac{2 T \chi(0,0)}{\Gamma} \underbrace{ \left( \int \frac{d\omega}{2\pi} W_\tau(\omega) \right)}_\tau \int dq \, \frac{q^3 e^{-2qd}}{(1+ q^2 \xi^2)^2} \nn
& \approx & \tau \left( \frac{ T^{(-2 + \eta)/z}}{\xi^4} \right) \ln\left( \frac{\xi}{d} \right) = \tau \left( \frac{T^{(2 + \eta )/z}}{c^4} \right) \ln\left( \frac{c}{d T^{1/z}} \right),
~ \text{ when } \omega_0 \tau \sim \Gamma \tau \sim T \tau \gg 1 \text{ and }  d \ll \xi
\label{eq:TFIMnoiseApp}
\eeq
Eq.~\eqref{eq:TFIMnoiseApp} completes the derivation of the noise at criticality for the quantum paramagnet to ferromagnet transition, quoted in Table~I in the main text.

\section{Numerical estimates of decoherence rate}
In this section, we provide additional details for the quoted decoherence rate of a probe spin localized nearby a monolayer CrI$_3$ sample. 
In CrI$_3$, the active degrees of freedom are $S = 3/2$ spins that are arranged on a hexagonal lattice \cite{mcguire2015}. 
Few-layered CrI$_3$ exhibits unusual magnetic phenomena at low temperatures, with magnetic ordering that depends on the number of layers \cite{mcguire2015,liu2019}. 
Here, we focus on monolayer CrI$_3$, which undergoes a ferro-magnetic transition at $T_\mathrm{c} = 45$ K \cite{gong2017,huang2017}. 
This thermal phase transition is an Ising transition with order parameter $S^{\mathrm{z}}$, and the spin-interaction Hamiltonian for CrI$_3$ conserves $S^{\mathrm{z}}$ \cite{Lado2017}. 
Thus, symmetries ensure that the Ising order parameter is conserved, and its diffusive dynamics near the thermal transition should be described by Model B, as discussed previously. 

To obtain a tentative lower-bound on the decoherence rate, we consider temperatures higher than $T_\mathrm{c}$; approaching the critical point by lowering $T$ will further increase the decoherence rate. 
This allows us to focus on the diffusive regime with relatively short spin-correlation lengths, such that the relevant parameters can be estimated more accurately from existing literature.
Further, as we will show explicitly below, in this regime the typical frequency scale $\omega_0 \sim D_s/d^2$ is in the THz regime for typical qubit-sample distance $d \sim 10$~nm, applicable to NV centers probe spins in diamond. 
Thus, the frequency-width of the noise spectrum $\mathcal{N}(\omega)$ is much broader than the range of frequencies over which the qubit probe is sensitive, i.e. where $W_\tau(\omega)$ is appreciably large, (see Fig.~\ref{fig:NcrossoverEcho}).
Consequently, we can approximate $\mathcal{N}(\omega)$ by $\mathcal{N}(\omega = 0)$, simplifying our analysis:
\beq\label{eq:CrI3_phase}
\langle \phi^2 \rangle = \int \frac{d\omega}{2\pi} W_\tau(\omega) \mathcal{N}(\omega) \approx  \mathcal{N}(0) \int \frac{d\omega}{2\pi} W_\tau(\omega) = \left( \frac{\gamma}{2\hbar} \right)^2 \tau \mathcal{N}(0)
\eeq
where we have used that the total integrated weight does not depend on the specific frequency filter function. 
Given the insensitivity to the precise frequency range considered, this enables us to directly make an estimation of \emph{spin-echo} $T_2^{\rm echo}$ time scale.
This is particularly useful, because, in experiments, one often prefers to work with spin echo pulse sequences as they efficiently cancel out external low frequency noise (such as shot-by-shot fluctuations), and enable longer coherence times for the probe spin.
Given the linear growth of the the decoherence phase (Eq.~\ref{eq:CrI3_phase}), the measured coherence profiles follows a simple exponential decay and thus the decoherence rate immediately informs us of the spin echo coherence time $T_2^{\rm echo}$: $2 \langle \phi^2 \rangle|_{\tau = T_2^{\rm echo}} = 1$.
Restoring all previously suppressed factors of $\hbar$ and $k_B$, we find:
\beq
\frac{1}{T_2^{\rm echo}} &=& 2\left( \frac{\gamma}{2\hbar} \right)^2 \mathcal{N}(0) = 2 \left( \frac{\gamma}{2\hbar} \right)^2 \left( \frac{g_s \mu_B \mu_0 S}{2a^2} \right)^2 \int_0^\infty dq \, q^3 e^{- 2 q d}  \left( \frac{2k_B T \chi_u}{D_s q^2} \right) \nn
&=& 2 \left( \frac{\gamma}{2\hbar} \right)^2  \frac{(g_s \mu_B \mu_0 S)^2}{16 \pi a^4 d^2} \left( \frac{\chi_u k_B T}{D_s} \right)
\eeq
We can estimate the uniform susceptibility as $\chi_u = \lim_{\omega \to 0} \lim_{q \to 0} \chi(\q,\omega) = \xi^2/J$.
Further, we assume that the temperature-dependence of the spin-diffusion constant $D_s$ appears only via the correlation length $\xi$.
Such an assumption is always true in the scaling limit where $\xi \sim (T_\mathrm{c} - T)^{-\nu}$, with $\nu = 1/2$ in mean-field theory.
Within mean-field theory, we can estimate $D_s \sim a^4 J/\xi^2 \hbar$; note that $\xi \approx a$ gives the expected high temperature limit ($T_\mathrm{c} \ll T \lesssim J$) of $D_s$ with a mean-free path $\sim a$ and an intrinsic interaction-induced scattering time $\sim \hbar/J$.
Using the aforementioned scaling assumption, we find that:
\beq
\frac{1}{T_2^{\rm echo}} = 2\left( \frac{\gamma}{2\hbar} \right)^2  \frac{(g_s \mu_B \mu_0 S)^2}{16 \pi a^4 d^2} \left( \frac{\hbar k_B T \xi^4}{J^2 a^4} \right)
\eeq

For CrI$_3$, we have a dominant interaction-scale $J = 2.2$ meV and in-plane lattice spacing $a = 0.687$ nm, while for NV qubits, $\gamma = g_\sigma \mu_B$ with the NV g-factor $g_\sigma = 2$ \cite{nvreview,schirhagl:2014a,casola:2018,pelliccione:2016,thiel:2019,nvsinglespin,Maze2011,Doherty2013,vool:2021}. 
Consider a temperature $T = 60$~K, so that the system is sufficiently away from the critical point such that $\xi \approx 2a$ (heuristically $\xi$ is of the lattice scale).
Then, for a sample-probe distance $d = 10$ nm, we find that $T_2^{\rm echo} \approx 5~\mathrm{\mu s}$, as quoted in the main text. 
Note that our estimate here if fairly conservative, approaching $T_\mathrm{c}$ to increase $\xi$ will enhance the noise several-fold.

\section{Beyond the Gaussian noise approximation}
In this section, we elaborate on the consequences of non-Gaussian noise, and provide a derivation of the scaling of non-Gaussian noise near the transition in terms of higher order correlation functions of the critical classical or quantum field $\varphi(\r,t)$. 
To this end, we start by noting that in the most general case (going beyond the Gaussian approximation), the qubit coherence as a function of time is given by $\langle \sigma_+(t) \rangle = \langle \sigma_+(0) \rangle e^{- \alpha(t) + i \theta(t)}$. 
The decay parameter $\alpha(t)$ and the phase rotation $\theta(t)$ are given by \cite{Viola16}:
\beq
\alpha(t) = \sum_{\ell=1}^\infty \frac{(-1)^\ell}{(2\ell)!} \Upsilon^{(2\ell)}(t), \text{ and }
\theta(t) = \sum_{\ell=1}^\infty \frac{(-1)^\ell}{(2\ell+1)!} \Upsilon^{(2\ell+1)}(t) \nn
\Upsilon^{(k)}(t) \equiv \int_0^\tau dt_1 \int_0^\tau dt_2...\int_0^\tau dt_k \, C^{(k)}(t_1,t_2,...,t_k)
\label{eq:NGnoise}
\eeq
In Eq.~(\ref{eq:NGnoise}), the k$^{th}$ order cumulant $C^{(k)}(t_1,t_2,...,t_k)$ is determined by the source field correlations $\langle B(t_1) B(t_2) .... B(t_j)\rangle$ with $j \leq k$. 
Focusing on noise generated by fluctuating magnetic fields, the relevant sources correspond to spin or current in the sample of interest.
Therefore, the k$^{th}$ order cumulant can be written as the convolution of the connected correlation functions in the sample and the appropriate kernel (obtained by solving Maxwell's equations) that propagate the field from the source to the qubit location (as described above when deriving the momentum filter function).
For insulators with spin-degrees of freedom, this kernel decays as $\sim q e^{-qd}$ [see Eq.~\eqref{eq:MaxwellKernel}], and the relevant correlation function is the connected $k$-point correlator  $\langle \varphi(\q_1,t_1)... \varphi(\q_k, t_k) \rangle$~\footnote{In this discussion, we restrict ourselves to cases where the order parameter is the coarse-grained magnetization.}. Indeed, the $k = 2$ case corresponds to the Gaussian noise discussed before; we now consider extensions to general $k$. 

To start, we note the cumulants or connected correlation functions with an odd number of fields determined the phase evolution of the decoherence, while the cumulants with an even number of fields determine its amplitude. 
For a ferromagnetic transition, this implies that the odd correlators are zero on the disordered side (where the $\varphi \rightarrow - \varphi$ symmetry is preserved), while they become non-zero on the ordered side as the symmetry spontaneously breaks (this can also be seen, for example, by expanding the field operator about the non-zero minima in the $\varphi^4$-theory, where odd terms appear beyond the quadratic approximation). 
Therefore, we expect the probe to pick up an additional phase once the symmetry is spontaneously broken; this can serve as a simple diagnostic of the sample's phase of matter.

We can also find the contribution of the higher order cumulants to the noise using scaling arguments. 
Let us consider the k$^{th}$ order term (with $k$ even), and study its contribution, $\alpha^{(k)}(t)$, to the decay dynamics, $\alpha(t) = \sum_{k=1}^\infty \alpha^{(2k)}(t)$. We note that, for a Ramsey sequence, this contribution can be written as:
\beq
\alpha^{(k)}(t) \sim \prod_{i=1}^{k} \int_0^\tau dt_i \left\langle \prod_{i=1}^{k} B(t_i) \right \rangle \sim \prod_{i=1}^{k} \int d\omega_i f_p(\omega_i)  \prod_{i=1}^{k} \int d^2 q_i \, q_i e^{- q_i d} \left \langle \prod_{i=1}^{k} \varphi(\q_i,\omega_i) \right \rangle
\eeq
where in the second line, $f_p(\omega_i) = \int_0^\tau dt \, e^{i \omega_i t}$ determines to the frequency filter function. 
More complicated control pulse sequences, denoted schematically by $y_p(t)$, can be taken into account by using $f_p(\omega_j) \equiv \int_0^\tau dt \, e^{i \omega_j t} y_p(t)$, as discussed in Ref.~\onlinecite{Viola16} and in analogy to our discussion about regarding the frequency filter function.
Since these do not make a difference to our scaling arguments, we restrict to a Ramsey sequence [$y_p(t) = 1$] for the following discussion. 

Let us first consider the case of a quantum critical point which is both Lorentz invariant ($z = 1$) and conformally invariant in $D = 2 + 1$ dimensions. 
In this case, conformal invariance dictates the following scaling relation for the multipoint correlation function of the order parameter field at $T = 0$:
\beq
\left \langle \prod_{i=1}^{k} \varphi(g \r_i, g t_i) \right \rangle = g^{- k \bar{\Delta}} \left \langle\prod_{i=1}^{k}  \varphi(\r_i, t_i) \right \rangle
\eeq
where $g$ denotes a re-scaling of spacetime, and $\bar{\Delta} = (D - 2 + \eta)/2$ is the scaling dimension of the field $\varphi(\r_i, t_i)$ \cite{ssbook} (we use $\bar{\Delta}$ for the scaling dimension to distinguish from the spectral gap $\Delta$).
Accordingly, in momentum space, we can write the the following scaling relation:
\beq
 \left \langle \prod_{i=1}^{k} \varphi(g\q_i,g\omega_i) \right\rangle =  g^{k(\bar{\Delta} - D)} \left\langle \prod_{i=1}^{k} \varphi(\q_i,\omega_i) \right\rangle 
 \label{eq:ScalingMP_Qspace}
\eeq
Rescaling the correlator with $g = 1/q_1$ enables us to write Eq.~\eqref{eq:ScalingMP_Qspace} as follows:
\begin{align}
\left \langle \prod_{i=1}^{k} \varphi(\q_i,\omega_i) \right \rangle &= q_1^{k(\bar{\Delta} - D)} \left\langle \prod_{i=1}^{k} \varphi\left(\frac{\q_i}{q_1},\frac{\omega_i}{q_1} \right) \right \rangle \\
&= q_1^{k(\bar{\Delta} - D) + D} \, \delta^{(D-1)}\left(\sum_{i=1}^{k} \q_i\right) \delta \left(\sum_{i=1}^{k} \omega_i\right) \, \left[ \left \langle \prod_{i=1}^{k} \varphi\left(\frac{\q_i}{q_1},\frac{\omega_i}{q_1} \right) \right \rangle \right]
 \label{eq:ScalingMP_Qspace2}
\end{align}
where we used spatial- and time-translation invariance to simplify the correlation function such that it depends only on $k-1$ momenta and frequency coordinates, $\q_k = - \sum_{i=1}^{k-1} \q_i$ with $\omega_k = - \sum_{i=1}^{k-1} \omega_i$. 
Since, in Eq.~\eqref{eq:ScalingMP_Qspace2}, the correlations inside square brackets only depend on dimensionless ratios, the dimensionful content of the multi-point correlator is simply set by the prefactors. 

Putting everything together, we can write the scaling form for the contribution to the noise $\alpha^{(k)}(t)$ due to $k$-point correlations at a Lorentz invariant quantum critical point ($z =1, \lambda = \lambda_c, ~ T = 0$):
\beq
\alpha^{(k)}(t) &\sim & \tau \left[\prod_{i=1}^{k} \int d^{(D-1)}q_i \, [q_i e^{- q_i d}] \right]  q_1^{k(\bar{\Delta} - D) + D} \, \delta^{(D-1)}\left(\sum_{i=1}^{k} \q_i\right) \times \tilde{\psi}_{sc}\left(\left \{\frac{q_i}{q_1}\right\}, \frac{1}{\tau q_1}\right) \nn
&=& \tau \times \frac{1}{d^{(k-1)(D-1)}} \times \frac{1}{d^{k}} \times \frac{1}{d^{k(\bar{\Delta} - D) + D}} \times \psi_{sc}\left( \frac{\tau}{d} \right) \nn 
&=& \frac{\tau}{d^{1 + k \bar{\Delta}}} \psi_{sc}\left( \frac{\tau}{d} \right) \xrightarrow{D = 2+1} \frac{\tau}{d^{1 + k (1 + \eta)/2}} \psi_{sc}\left( \frac{\tau}{d} \right) 
\eeq
where the delta function on the sum of frequencies contributes a factor of $\tau$ due to the frequency integrals,
$\tilde{\psi}_{sc}$ and $\psi_{sc}$ are dimensionless scaling functions, and we have used the fact that we are working in two spatial dimensions. 

So far we have studied a very particular case, namely a quantum critical point with Lorentz and conformal invariance.
However, deviations away from this point will modify the scaling form above. 
For example: 
(i) the critical point may not have Lorentz invariance (i.e., $z \neq 1$);
(ii) the sample may not be at the critical point (i.e., $\lambda \neq \lambda_c$, and thus the spectral gap $\Delta \neq 0$);
and (iii), the temperature will not be zero (i.e., experiments always work at small but non-zero temperatures $T > 0$).
Nevertheless, near a quantum critical point, we expect the multi-point correlation function to be described by a scaling form that is governed by the quantum critical point. 
Therefore, we generalize the scaling form [Eq.~\eqref{eq:ScalingMP_Qspace2}] by taking $D = 2 + 1 \to 2 + z$, such that $\bar{\Delta} = (2 + z - 2 +\eta)/2 = (z + \eta)/2$, and introducing additional scaling parameters $\omega_i/T$ and $\Delta/T$:
\beq
\left\langle \prod_{i=1}^{k} \varphi(\q_i,\omega_i) \right\rangle = q_1^{k(\bar{\Delta} - 2 - z) + (2+z)} \, \delta^{(2)}\left(\sum_{i=1}^{k} \q_i\right) \delta \left(\sum_{i=1}^{k} \omega_i\right)  \, \tilde{\Psi}_{sc}\left( \left \{\frac{c q_i}{T^{1/z}}\right\}, \left\{ \frac{\omega_i}{T} \right \}, \frac{\Delta}{T} \right)
\label{eq:kPtScalingCFT}
\eeq
where $\tilde{\Psi}_{sc}$ is a dimensionless scaling function.
Accordingly, the scaling form for the contribution to noise $\alpha^{(k)}(t)$ due to $k$-point correlations is given by:
\beq
\langle \phi^2 \rangle_k \sim T^{k(2 + \eta - z)/(2z)} \Psi_{sc}\left( \tau T, \frac{d T^{1/z}}{c}, \frac{\Delta}{T} \right)
\eeq
We can check that setting $k = 2$ leads us to the familiar expression for Gaussian noise previously derived in Eq.~\eqref{eq:IsingScaling2pt}, while the relativistic conformal limit $(z =1, T = 0, \Delta = 0)$ reproduces Eq.~\eqref{eq:kPtScalingCFT}.
Note that away from criticality in presence of a spectral gap $\Delta$ or a correlation length $\xi \sim 1/\Delta$ (for $z = 1$ theories), short-range correlations imply that there is a scaling factor $(\xi/d)^{3k -2}$ which heavily suppresses non-Gaussian ($k \geq 4$) contributions to decay of coherence at large $d$. 
This is easier to intuit in real space, where each dipolar contribution gives a factor of $d^{-3}$, and spatial translation invariance gives $d^2$, leading to an overall scaling of noise as  $d^{-3k+2}$. 
However, this is not necessarily true as we approach criticality (i.e., $\Delta \to 0$ or $\xi \to \infty$ diverges). 
Nevertheless, in realistic systems, presence of domains restrict the divergence of $\xi$. 
Furthermore, we still expect the higher order contributions to noise to be suppressed by a power-law of $T$ as low temperatures even when $\lambda = \lambda_\mathrm{c}$, simply because there is low density of fluctuations at such low temperatures. 
Taken together, these suggest that the Gaussian approximation can provide a good description of decoherence in the vicinity of criticality, provided the sample-probe distance $d$ is not very small. 
Finally, we note that dynamical decoupling protocols that use suitably designed dynamically decoupling pulse sequences can extract non-Gaussian contributions, as has been shown both theoretically \cite{AS11,Viola16} and experimentally for superconducting qubits \cite{Sung19}.

\section{Differentiating bulk and surface critical phenomena}

Throughout the main text, we have assumed that the sample of interest undergoes a single phase transition as we tune the experimental parameters.
However, for thick material samples, the surface and bulk degrees of freedom can exhibit widely different behaviors; most notably, exhibiting distinct critical phenomena~\cite{binder1983phase, cardy1996scaling, Satcher}.
Crucially, both critical phenomena will induce different critical fluctuations, affecting the qubit's decoherence dynamics differently.
Understanding how each type of fluctuation affects the qubit's dynamics is highly dependent on the details of the material and the nature of the two critical behaviors.
However, some important considerations can be made.

Most importantly, one must note that the sample-probe distance determines the relative contributions of surface and bulk fluctuations to the qubit's decoherence; thus, it offers a controllable parameter capable of isolating and characterizing each critical phenomena separately.
Indeed, although in the main text we have focused on the role of $d$ in selecting a particular momentum along the surface direction, the evanescent nature of the momentum filter function ($e^{-2qd}$ in Eq.~\ref{eq:Wdq}), also determines the depth of the fluctuations that the spin-probe qubit couples to.

To highlight this piece of physics, it is important to consider the two relevant lengthscales in the problem: the width of the surface layer $W$ and the sample-probe distance $d$.
When $d\lesssim W$, the probe-spin qubit can only couple to the degrees of freedom at the surface since the fluctuations are dominated by signatures of the surface criticality.
We note that this highlights the flexibility of our proposal for directly studying surface criticality by placing the qubit probe very close to the material sample.
When $d\gg W$, the spin-qubit probe becomes sensitive to a volume of the sample that (mostly) includes bulk degrees of freedom.
As a result, most of the local fluctuating fields (and thus the decoherence dynamics) are expected to arise from the bulk fluctuations and thus one should be able to characterize that particular criticality.

While these considerations provide a qualitative understanding of the regimes under which we expect to be sensitive to the bulk or the surface criticality, other effects need to be taken into account for a full description of the decoherence dynamics.
Indeed, if one aims to quantitatively extract the criticality of the bulk (or the surface) from the decoherence signal, one must first generalize our formalism to include the effect of correlations of the fluctuations along the direction normal to the sample's surface.
Subsequently, the presented scaling analysis must be modified to account for the two nearby critical phenomena, incorporating the coupling strength between the spin-probe and the fluctuating degrees of freedom, as well as the nature and position of the different critical points.
We leave this detailed analysis for future work.

Finally, let us note that there are settings where we expect the surface criticality to not play a role in the decoherence signal.
For example, surface sensitive probes such as scanning tunneling microscopy (STM) and angle resolved photoemission spectroscopy (ARPES) have long been used to study the        bulk properties of materials \cite{Chen_2021,Sobota_He_Shen_2021}, suggesting that a judicious choice of materials and their preparation can enable a direct investigation into the materials properties.

\bibliography{references}